\documentclass[aps,prd,superscriptaddress,reprint,nofootinbib]{revtex4-1}
\usepackage{graphicx}
\usepackage{dcolumn}
\usepackage{bm}
\usepackage{lipsum}
\usepackage{amsmath,amssymb}
\usepackage{dsfont}
\usepackage{float}
\usepackage{orcidlink}
\usepackage{array}
\makeatletter
\let\newfloat\newfloat@ltx
\makeatother
\usepackage{algorithm}
\usepackage{algpseudocode}
\usepackage{multirow}
\usepackage{colortbl}
\usepackage{booktabs}
\usepackage{stmaryrd}
\usepackage{minted}
\usepackage{empheq}
\usepackage[most]{tcolorbox} 
\usepackage{makecell}
\usepackage{xcolor}  
\newcolumntype{C}{>{$}c<{$}}

\AtBeginDocument{%
  \heavyrulewidth=.08em
  \lightrulewidth=.05em
  \cmidrulewidth=.03em
  \belowrulesep=.65ex
  \belowbottomsep=0pt
  \aboverulesep=.4ex
  \abovetopsep=0pt
  \cmidrulesep=\doublerulesep
  \cmidrulekern=.5em
  \defaultaddspace=.5em
}

\usepackage{verbatim}
\usepackage[left=2cm, right=2cm, top=2cm]{geometry}

\usepackage[normalem]{ulem}
\usepackage[utf8]{inputenc}
\usepackage{epigraph}
\usepackage{siunitx}
\usepackage[english]{babel}
\usepackage{listings}
\usepackage{graphicx}
\usepackage[utf8]{inputenc}
\usepackage{listings}
\usepackage{matlab-prettifier}
\usepackage{xcolor}
\usepackage{amsfonts}
\usepackage{amsmath}
\usepackage{amssymb}
\usepackage{amsthm}
\newcolumntype{P}[1]{>{\centering\arraybackslash}p{#1}}
\newcolumntype{M}[1]{>{\centering\arraybackslash}m{#1}}

\usepackage{braket}
\usepackage{xcolor}
\usepackage{tcolorbox} 
\usepackage{color}

\newcommand{\be}{\begin{equation}}
\newcommand{\ee}{\end{equation}}
\newcommand{\delay}{\mathbf{D}}

\hypersetup{
    colorlinks=true,
    linkcolor=blue,
    filecolor=magenta,      
    urlcolor=cyan,
    pdftitle={Overleaf Example},
    pdfpagemode=FullScreen,
    citecolor=blue
    }

\begin{document}

\title{The impact of missing data on the construction of LISA Time Delay Interferometry Michelson variables}

\author{Ollie~Burke\,\orcidlink{0000-0003-2393-209X}}\email{contact: ollie.burke@glasgow.ac.uk}
\affiliation{School of Physics and Astronomy, University of Glasgow, Glasgow G12 8QQ, UK}
\author{Martina~Muratore\,\orcidlink{0000-0002-9630-5698}}\email{contact: martina.muratore@aei.mpg.de}
\affiliation{Max Planck Institute for Gravitational Physics (Albert Einstein Institute), D-14476 Potsdam, Germany}
\author{Graham Woan\, \orcidlink{0000-0003-0381-0394)}}
\affiliation{School of Physics and Astronomy, University of Glasgow, Glasgow G12 8QQ, UK}

\begin{abstract}

We investigate the impact of missing input data on the construction of second-generation Time Delay Interferometry (TDI) variables, which enable data analysis for the Laser Interferometer Space Antenna (LISA). TDI relies on the introduction of precise time delays into the raw interferometric data streams before they are combined to suppress otherwise dominant laser phase noise. We show that a single missing sample, corresponding to 0.25\,s of data, will result in an effective data gap of approximately 90\,s in the second-generation TDI output if further measures are not taken. This additional gap is largely independent of the initial gap duration, but increases linearly with the order of the fractional-delay filter used for the computations. For a realistic gap scenario, incorporating both planned and unplanned data interruptions consistent with a target duty cycle of $\sim 84$\%, we find that frequent, short-duration gaps (e.g., a total of 1000 per year, each of which have short durations $\sim $ 100 s) could result in an additional loss in the TDI variables of about one day per year corresponding to a $\sim 0.3$\% reduction in duty cycle. This amounts to a loss of approximately one day of LISA data suitable for the global-fit per year. 
\end{abstract}
\date{\today}
\maketitle

\section{Introduction}

The Laser Interferometer Space Antenna (LISA) is a European Space Agency led space-based gravitational wave (GW) observatory with the goal of observing astrophysical gravitational waves frequencies between $0.1$\,mHz and $1$\,Hz ~\cite{LISA:2024hlh,LISA:2017pwj}. Unlike ground-based detectors, LISA is expected to be signal dominated, observing a multitude of GWs that overlap in both time and frequency~\cite{Colpi:2019yzd, berry2019uniquepotentialextrememassratio}.
Current GW inference techniques for LISA are not designed to operate directly on the raw phasemeter data. Instead, they take pre-processed data from an initial noise reduction pipeline that transforms the raw phasemeter measurements into calibrated data streams suitable for analysis~\cite{Tinto2014}. This is due to the dominance of laser frequency noise---arising both from intrinsic laser instability and arm-length variations---which exceeds the (GW) astrophysical signal by eight to nine  orders of magnitude in amplitude. In contrast, ground-based GW detectors like LIGO, VIRGO and KAGRA are able to suppress laser noise almost perfectly by constructing near-equal arm interferometers~\cite{Aasi2015,Abbott2016a}. LISA operates as a formation-flying constellation in space, with three spacecraft arranged in an approximately equilateral triangle that trails Earth in a heliocentric orbit inclined by 60 degrees to the ecliptic~\cite{colpi2024lisadefinitionstudyreport}. While the nominal arm lengths are $\sim2.5 \times 10^6$\,km, they are neither equal nor constant over the mission lifetime. As a result, LISA cannot adopt a traditional equal-arm Michelson interferometer design. To address this, LISA employs Time-Delay Interferometry (TDI), a technique that synthesizes virtual equal-arm interferometers by appropriately time-shifting and linearly combining measurements taken along different arms. This process effectively suppresses the laser frequency noise to levels compatible with GW science objectives~\cite{LISA2017,Hartwig2021,Staab2023,Reinhardt2024,Bayle2019,Tinto2020}.

\begin{figure}[ht]
\includegraphics[width=\linewidth]{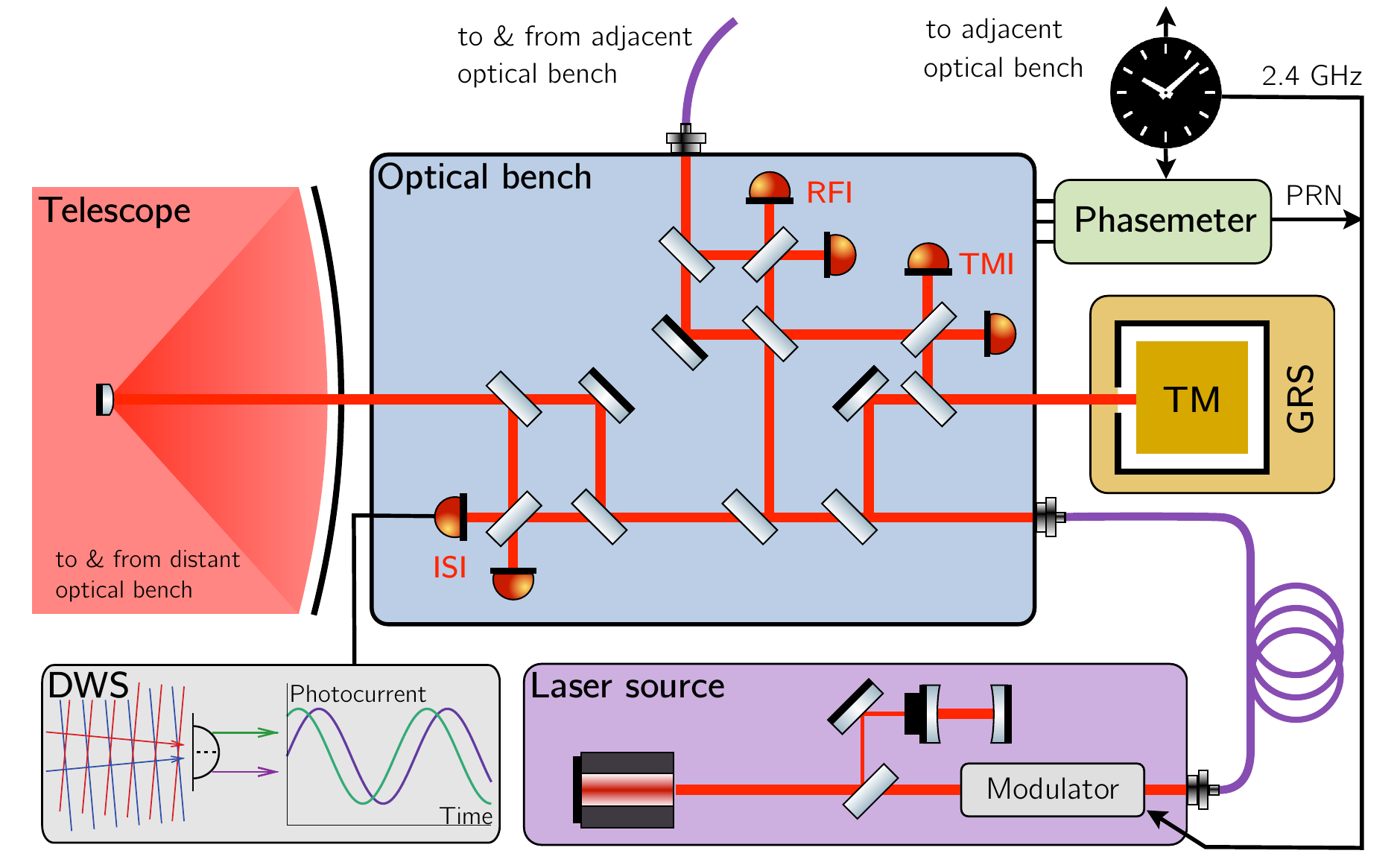}
\caption{\label{clock_picture} 
        LISA's measurement system, adapted from \cite{Bayle_2023} with support from O.~Hartwig. The Figure shows how each spacecraft uses three main interferometers (IFOs) to detect GWs. The Interspacecraft IFO captures the GW signal, buried within much louder noise, primarily from laser frequency fluctuations. The other two IFOs, the Reference and Test-mass IFOs, help remove noise due to spacecraft motion. Additionally, clock sidebands and pseudorandom noise codes correct for timing and laser noise, while the differential wavefront sensing measures beam angles to minimize angular jitter effects.}
\end{figure}

Each LISA spacecraft (S/C) hosts two Moving Optical Sub Assemblies (MOSAs), and each of these carries an optical bench equipped with three distinct interferometers (IFOs): (i) an inter-satellite IFO (ISI), (ii) a test-mass IFO (TMI), and (iii) a reference IFO (RFI) \cite{LISA:2024hlh}. The test-mass IFO measures sub-nanometre fluctuations in the displacement between the optical bench (OB) and its associated free-falling test mass which provides a local inertial reference. The inter-satellite IFO interferes the incoming light from a remote spacecraft with locally generated laser light, measuring the relative optical path length between distant OB and local OB. Finally, the reference interferometer measures the relative displacement between the two OBs hosted on the same S/C. These give a total of 18 interferometric phasemeter measurements. In total however, the full raw phasemeter data has 66 channels, including clock sidebands and measured pseudoranges (MPRS) comprising the inter-spacecraft distance plus clock offset and ranging noises, that are needed to fully correct for timing and laser noise. Additional differential wavefront sensors measure the laser beam angles, and are used to minimize the effects of angular jitter \cite{Hartig2022,Paczkowski2022,Hartig_2025}. All these measurements are sampled at 4\,Hz before transmission to Earth, and are summarised in Fig.~\ref{clock_picture}.

Given the complexity of the LISA instrument and operations \cite{LISA_PerfDoc_ESA_2022}, data artifacts are to be expected \cite{Castelli:2024sdb,Burke:2025bun}.  Here, we will focus on the impact of missing data (that is, data gaps) potentially due to corrupted data that is removed having been rendered unsuitable for further analysis. The gap pattern considered in later sections of our study represent plausible scenarios that incorporate both scheduled short-term interruptions and unscheduled long-term data gaps.

Scheduled short gaps are expected when scientific measurements are temporarily disrupted by operational activities. For instance, these may occur during the periodic repositioning of the high-gain antenna to maintain Earth communication, or during adjustments made by the point-ahead angle mechanism (PAAM) to correctly align laser links. Although not strictly periodic, such gaps are anticipated to recur with near-regular frequency. Their duration may vary: antenna-related gaps could last a few hours and occur approximately every two weeks (medium duration gap), while PAAM-related gaps are expected to be shorter—--typically under 100\,s—--occurring a few times per day. 

In contrast, unscheduled long-duration interruptions are random and typically arise from unexpected anomalies. These may include events such as system failures requiring recovery procedures, similar to those encountered in LISA Pathfinder, or impacts from micrometeoroids. In the scenario adopted here, we assume several such events may occur annually, each lasting anywhere from one to three days. 

Routine maintenance, as well as periods of partial or complete data loss are also expected. The nature and frequency of planned gaps will depend on the finalized design and operational strategy of the spacecraft and instruments.

To the author's knowledge, all existing studies on the impact of data gaps~\cite{Burke:2025bun, Mao:2024jad, Castelli:2024sdb, Dey:2021dem, Baghi:2019eqo, Carre:2010ra, Wang:2024ovi} have focused on signal recovery from post-processed gapped TDI observables, i.e., after the suppression of laser frequency noise, and none have examined the effects of missing phasemeter data on TDI itself. The goal of this work is to investigate how data gaps present in the scientific telemetry variables propagate through the initial noise-reduction pipeline and affect the construction of the final TDI observables. In particular, we aim to quantify the `augmentation' of gaps: the increase in missing or invalid samples caused by the propagation of data gaps through intermediate processing steps. Understanding this compounding data loss is essential for ensuring that downstream stages of the LISA analysis pipeline, including parameter estimation and source population inference, are based on maximally preserved and well-characterised datasets. The global-fit for LISA aims to jointly recover both gravitational-wave sources and the underlying noise properties~\cite{Littenberg:2023xpl,Katz:2024oqg,deng2025modular,PhysRevD.110.024005}. Accurate noise modelling is crucial since likelihood functions such as the Whittle likelihood~\cite{whittle:1957} rely on assumptions of stationary, Gaussian, and circulant noise. Under these conditions, the Fourier transform diagonalizes the noise covariance, enabling efficient likelihood evaluation. However, real LISA data include effects like data gaps~\cite{Talbot:2021igi,Edy:2021par} and non-circulant (Toeplitz-like) noise, making the Whittle likelihood only an approximation. Mischaracterising the noise can bias parameter estimation~\cite{Burke:2025bun}, so it is essential to carefully track and understand noise contributions from raw spacecraft measurements through to TDI variables and the global-fit likelihood. The results presented here aim to inform future strategies for telemetry management and  gap mitigation.

This paper is organised as follows. In Sec.~(\ref{sec:LISA_post_processed_variables}), we give a brief discussion of TDI (\ref{subsec:TDI_variables}) and Lagrange-interpolation methods used to compute fractional delays (\ref{subsec:fractional_delays}). Section (\ref{sec:missing_data_section}) contains our main analytical results. Specifically, in subsection (\ref{subsec:telemetry_missing_data}) we discuss our analytical treatment of missing-data and gaps; and we compute the widening of the gaps in the intermediary variables in Sec.(\ref{subsec:intermediary_variables_gap_augmentation}), and in the TDI variables in Sec.(\ref{subsec:gap_augmentation_TDI_variables}). Our analytical results are supported via appendices (\ref{app:gap_augmentation_eta}-- \ref{app:multiple_gaps_telemetry_variables}).  Our results section is given in Sec.(\ref{sec:Results}). In Sec.(\ref{subsec:results_verification}) we numerically verify our analytical results in Fig.~\ref{fig:TDI_Variables_X_numerical_plot}. To conclude, we then perform full LISA-like simulations in Sections (\ref{subsec:results_full_gaps_telemetry}) and (\ref{subsec:results_full_scale_TDI_simulation_gaps}) to understand (1) the duty cycle of the LISA mission and (2) the overall loss of data in a four-year LISA-like simulation due to TDI post processing on missing data. 
Finally, throughout our work, we will follow the conventions of PyTDI and LISA Instrument \cite{bayle2023lisainstrument,Staab2022PyTDI}.

\section{Time Delay Interferometry}\label{sec:LISA_post_processed_variables}

To extract GW signals from LISA's interferometric data, it is essential to construct intermediate observables that suppress the most dominant noise sources (e.g.spacecraft motion (jitter) and laser frequency noise) to ultimately reduce the number of effective lasers from six to three. This is achieved through the derivation of intermediary variables, starting from the basic interferometric measurements collected onboard the spacecraft. Since there is already a plethora of material in the literature regarding TDI, we will report only a brief summary that is relevant for overall scope of this work. On this subject, we will not describe all instrumental noise sources which are both well-characterized and measurable. This includes, for example, the optical tilt-to-length coupling induced by spacecraft motion and clock noise (see \cite{Hartig_2022}, \cite{PhysRevApplied.14.014030}, and \cite{PhysRevD.103.123027}). Residual noises refer to additional sources of noise that cannot be mitigated by the initial noise reduction pipeline. We can group these residual noises into two main types: a test-mass (TM) acceleration noise and a general optical metrology system (OMS) noise contribution associated with each individual inter-spacecraft link. We refer the reader to \cite{nam2022tdi} for extensions involving multiple OMS noise components)

We will consider for our analysis only the three IFOs measurements \cite{Muratore:2021rwq} (See Fig.\ref{clock_picture} for reference) per optical bench
\begin{itemize}
    \item $\text{isi}_{ij}$ — science IFO measuring the optical path between the laser hosted in the optical bench of the distance S/C $j$ and the local laser $i$,
    \item $\text{rfi}_{ij}$ — reference IFOs interferes the local laser $i$ and the laser in the adjacent optical bench $j$
    \item $\text{tmi}_{ij}$ — test mass IFO interferes the laser in the adjacent optical bench $j$ with the local laser $i$ once this has bunched off the test mass.
\end{itemize}

Note that here, the indices \( ij \) refer to signals emitted by S/C \( j \) and received by S/C \( i \). For instance, \( \text{isi}_{12} \) captures the signal transmitted from S/C 2 and S/C by craft 1.

\subsection{Construction of Time-Delay Interferometry Variables}\label{subsec:TDI_variables}
The first step in post-processing is the formation of the first class of intermediary \(\xi_{ij}\) variables, which are designed to suppress spacecraft jitter. These are computed as follows:
\begin{align}\label{eq:xi_variables}
\nonumber
\xi_{ij} = \text{isi}_{ij} &+ \frac{1}{2}(\text{rfi}_{ij} - \text{tmi}_{ij}) \\
                           &+ \frac{1}{2} \dot{\mathbf{D}}_{ij}(\text{rfi}_{ji} - \text{tmi}_{ji}),
\end{align}
where \(\dot{\delay}_{ij} \) is the Doppler-delay operator associated with the light travel time from S/C \( j \) to \( i \). 

For a generic time series \( x(t) \), the Doppler-delay is applied as \cite{PhysRevD.65.082003}
\begin{align}
\dot{\mathbf{D}}_{ij}x(t) &= (1 - \dot{d}_{ij})\delay_{ij}x(t) \nonumber \\ & = (1 - \dot{d}_{ij})x(t - d_{ij}(t)) \label{eq:doppler_delay} 
\end{align}
with \( d_{ij}(t) \) denoting the (possibly time-varying) light travel time and $\dot{d}_{ij}$ the time-derivative of the light travel time between the two S/C $j$ to $i$ with arm-length $L_{ij}$ with $d_{ij}(t) = L_{ij}(t)/c$. In our analysis, we are working in variables expressed in total frequency, rather than phase resulting in the use of Doppler-based delay operators \cite{PhysRevD.105.122008}. 

Next, one constructs the second class of intermediary variables, \(\eta_{ij}\), which serve to reduce the full system to three effective, independent laser noises: 
\begin{subequations}\label{eq:etas_variables}
\begin{align}
\eta_{12} &= \xi_{12} + \frac{1}{2} \dot{\mathbf{D}}_{12}(\text{rfi}_{21} - \text{rfi}_{23}), \\
\eta_{13} &= \xi_{13} + \frac{1}{2}(\text{rfi}_{12} - \text{rfi}_{13}).
\end{align}
\end{subequations}
with four other variables $\eta_{21}, \eta_{23}$ and $\eta_{31},\eta_{32}$ given by cyclic index permutations of the equations in \eqref{eq:etas_variables}. 

To simplify the notation of sequential delays, we define chained delay operators \cite{PhysRevD.65.082003}:
\begin{equation}
\dot{\delay}_{i_1i_2\cdots i_n} = \dot{\delay}_{i_1i_2}\dot{\delay}_{i_2i_3}\cdots\dot{\delay}_{i_{n-1}i_n},
\end{equation}
which recursively apply delays. For instance,
\begin{widetext}
\begin{align*}
\dot{\mathbf{D}}_{ijk}x(t) = \dot{\mathbf{D}}_{ij}\dot{\mathbf{D}}_{jk}x(t) = [1 - \dot{d}_{ij}(t)][1 - \dot{L}_{jk}(t - d_{ij}(t))]x\big(t - d_{ij}(t) - L_{jk}(t - d_{ij}(t))\big),
\end{align*}
\end{widetext}
demonstrating the non-commutative nature of Doppler-delay operators when arm lengths vary over time. 

Using the \(\eta_{ij}\) variables, one can construct Michelson-like TDI combinations that cancel laser frequency noise. First- and second-generation TDI observables, which incorporate increasingly longer sequences of delays to accommodate time-dependent arm lengths, are given in factorized form:
\begin{equation}
\begin{aligned}
\mathbf{X}_{1} = (1 - \dot{\delay}_{131})(\eta_{12} + \dot{\delay}_{12}\eta_{21}) \ - \\ (1 - \dot{\delay}_{121})(\eta_{13} + \dot{\delay}_{13}\eta_{31})\label{eq:1st_gen_TDI_factorised}
\end{aligned}
\end{equation}
\begin{equation}
\begin{aligned}
\mathbf{X}_{2} = (1 - \dot{\delay}_{131} - \dot{\delay}_{13121} + \dot{\delay}_{1213131})(\eta_{12} + \dot{\delay}_{12}\eta_{21}) -\\ (1 - \dot{\delay}_{121} - \dot{\delay}_{12131} + \dot{\delay}_{1312121})(\eta_{13} + \dot{\delay}_{13}\eta_{31}).\label{eq:2nd_gen_TDI_factorised}
\end{aligned}
\end{equation}
The other TDI observables $\{\mathbf{Y}_{1},\mathbf{Y}_{2}\}$ and $\{\mathbf{Z}_{1},\mathbf{Z}_{2}\}$ are computed through cyclic permutation of the indices. The TDI observables form the backbone of LISA data analysis, enabling the extraction of GW signals through canceling the otherwise dominant laser noise. Many more TDI combinations can be built \cite{Muratore:2020mdf,Vallisneri:2005ji}, here we will only discuss the Michelson TDI because these are the current baseline used for LISA data analysis \cite{LDCSA,LDCS}.

\subsection{Fractional Delays}\label{subsec:fractional_delays}

Let $x(t)$ be a generic and continuous time-domain function. The discrete analogue of $x(t)$ is given by an evenly sampled sequence $x[n]$, with time-stamps $t_{n} = t_0 + n\Delta t$ with $t_0$ the initial starting time and $\Delta t = t_{n+1} - t_{n}$ the sampling interval. We define the observation time $T_{\text{obs}} = K\Delta t$ for $n \in [0,1,\ldots,K-1]$. 

To ease notation in this section, we will only consider the usual delay operator $\delay$ rather than the Doppler-delay $\dot{\delay}$ since the term $\dot{d}_{ij}$ will vanish for our configuration as discussed later. Our results are easily extended for more general LISA-based orbit configurations, which we will leave for future work. 

In general, due to the discrete nature of our time-series, the quantity $x(t - d_{ij}(t))$ will not exist as a discrete sample and must be approximated via interpolation schemes. The most common approach used throughout the literature is \emph{Lagrange Interpolation}~\cite{PhysRevD.109.043040}. Lagrange interpolation has a maximally flat transfer function at $f = 0$, which implies that this interpolating method is very accurate for low frequencies where LISA will operate. A more efficient, and numerically more stable interpolant was recently developed in~\cite{Staab:2024nyo}, which avoids (1) pathological behaviour at higher frequencies and (2) elevated number of terms (coefficients) in the interpolation scheme required to accurately infer the delay. The interpolating method found in~\cite{Staab:2024nyo} could easily be incorporated into this work, but will not change the results considerably. 

Now, suppose the delay $d_{ij}(t) \in \mathbb{R}^{+}$. Then following \cite{Hartwig:2021dlc}, we can always make the split
\begin{equation}
d_{ij}(t) = d_{ij}^{0}(t) + d_{ij}^{\epsilon}(t)\,,
\end{equation}
for integer part $d_{ij}^{0}(t)$ and fractional part $d_{ij}^{\epsilon}(t)$. In our discrete notations, $d^{0}(t) \approx d^{0}_{ij}[n]$ with integer shift $N^0_{ij}\in\mathbb{N}$ and fractional shift $N^{\epsilon}_{ij}[n]\in [0,1)$ given by
\begin{align}
N^0_{ij}[n] &= \lfloor d_{ij}[n]/\Delta t\rfloor = \lfloor L_{ij}[n]/c\Delta t\rfloor \\
N^{\epsilon}_{ij}[n] &= d_{ij}[n]/\Delta t - N^0_{ij}[n]
\end{align}

Notice that the integer part $N^0_{ij}[n]$ can be (1) time-varying due to the non-constant arm-lengths and (2) different (but constant) between individual links due to the arm-lengths being non-equal. Note that if we are to assume the individual links to be equal and non time-varying, resulting in a static orbital configuration, then $N^0_{ij}[n] = N$, i.e., a constant integer delay that is common across all arm-links. 

In terms of continuous IFO measurements, we have
\begin{align}
\delay_{ij}\text{ifo}_{ij}(t) &= \text{ifo}_{ij}(t - d_{ij}(t)) \nonumber \\ &= \text{ifo}_{ij}(t - d_{ij}^0(t) - d_{ij}^{\epsilon}(t)) 
\end{align}
which, in discrete notations, is given by
\begin{align}
\delay_{ij} x(t) &\approx (\delay_{ij}x)[n] \nonumber \\
&\approx \sum_{k=-p + 1}^{p}\text{ifo}_{ij}[n - N^0_{ij}[n] + k]\ell_{k}(-N^{\epsilon}_{ij}[n])\label{eq:approx_delay_x}
\end{align}
with $p = (\mathcal{K} + 1)/2$, for $\mathcal{K}$ the order of the Lagrange interpolant and $\ell_{r}(x)$ the Lagrange basis functions defined by 
\begin{equation}
\ell_r(x) = \prod_{\substack{j=-p+1}}^{p} \frac{x - j}{r - j}\,, \quad  j \neq r.
\end{equation}

The expression of Eq.\,\eqref{eq:approx_delay_x} is used to approximate the delay of a time-series at a particular (fractional) point. First and second-generation TDI variables defined by Eqs. \eqref{eq:1st_gen_TDI_factorised} and \eqref{eq:2nd_gen_TDI_factorised} feature \emph{nested} delays. Following \cite{Hartwig2021}, for simplicity, let $A_{1}$,$A_{2}$,\ldots $A_{n}$ represent potential S/C index pairings. Notice that
\begin{align}
\delay_{A_1A_2\dots A_{n}}x(t) &= \delay_{A_1}\delay_{A_2}\dots \delay_{A_n}x(t) \nonumber \\ & =x(t - d_{\text{total}}), \nonumber
\end{align}
with total time-delay 
\begin{equation*}
\begin{aligned}
d_{\text{total}}(t) &= d_1(t) + \delay_{A_1}d_{2}(t) + \dots + \delay_{A_{1}\dots A_{n-1}}d_{n}(t))\,,
\end{aligned}
\end{equation*}
that can be computed recursively in the following manner
\begin{align}\label{eq:d_total_nested}
d_{\text{total}}(t) &= \sum_{k=1}^{n}z_{k}(t) ,
\end{align}
with
\begin{align}
z_{1}(t) &= d_1(t), \\
z_{q}(t)&= d_{q}\left(t - \sum_{k=1}^{q-1}z_k(t)\right)\,. \label{eq:nested_delays_z_equations}
\end{align}
Equation \eqref{eq:nested_delays_z_equations} is then computed using Lagrange Interpolation. In the general orbital case where the LISA-arms are neither equal or constant, the delays are smooth functions with very low frequency content. For this reason, the recursive computation defined via Eq.\,\eqref{eq:nested_delays_z_equations} can be computed with very low interpolation orders that still results in highly accurate interpolated measurements \cite{Hartwig:2022yqw}.

In order to provide a simple analytic treatment that describes the impact of data gaps on TDI computations, we will consider a few simplifications to the orbital motion of LISA. We will assume that the arm-lengths of LISA are equal (and set to the average arm length of  $\bar{L} = L_{ij} = 2.5\cdot 10^{9}$\,m) and constant  ($\dot{L}_{ij} = 0$). We will neglect the effect of having three clocks with different time offsets and drifts, allowing us to discard the clock sideband measurements from the analysis. In the conclusion we will discuss how this choice could impact our results. Overall, we do not expect our results to change considerably for more complex orbits.

To conclude, for the remainder of the paper we will consider fractional delays acting on the time-series that can be approximated by
\begin{align}
\delay_{A_1}\dots \delay_{A_m}x(t) &\approx (\delay_{A_1}\dots \delay_{A_m}x)[n]\nonumber\\
&\approx \sum_{k = -p + 1}^{p}x(n - N + k)\ell_{k}(-\epsilon) ,
\end{align}
with
\begin{align*}
N &= \lfloor m\cdot \overline{d}\rfloor\,, \quad \text{Integer part} \\
\epsilon &= m\cdot \overline{d} - N \,, \quad  \text{Fractional part} \\
\end{align*}
for $m$ the number of delay operators and $\overline{d}$ the average light travel ($\overline{d}$ = 8.34\, s \cite{Martens:2021phh}) time defined between any two-links. 

\section{Result of Missing Data on TDI variables}\label{sec:missing_data_section}
\subsection{Loss of Scientific Telemetry}\label{subsec:telemetry_missing_data}
In the context of GW parameter estimation and general data analysis, it is reasonable to assume that missing data can be reflected via imputation with zeros~\cite{Burke:2025bun, Wang:2024ovi}. This amounts to a multiplication of the original time-series with a gating function, zero for $t \in T_{\text{gap}}$ and one otherwise. From a parameter estimation point of view, it has been shown that the ``windowing" method returns identical statistical quantities (likelihoods, signal-to-noise ratios, information matrices etc.) to the masking methods in either the time or frequency domain (more details can be found in ref.\cite{Burke:2025bun}). After TDI has been performed, then the \textit{global-fit} analyses should, in theory, be invariant to whether you apply windows or masks to your data stream, provided that the likelihood is consistent with the noise generation process. 

For the \textit{noise mitigation pipeline} this is not the case. We observed that treating missing-data in telemetry variables with zeros results in failure to cancel laser noise at the boundaries of the gaps. Imputing the scientific telemetry data with zero-values is particularly dangerous if there are non-trivial trends to the data stream since this would lead to spurious noise artifacts in the final TDI streams, potentially leading to pathological parameter inference during the global-fit scheme. A much cleaner and more elegant method is to assume complete ignorance of missing data -- via imputation with unrepresentable and undefined floating points -- \texttt{NaNs}\footnote{Strictly speaking, missing data of this form are represented by ``quiet" \texttt{NaNs}: missing-data that does not cause code to fail execution on detection of a \texttt{NaN}. As such, these quiet \texttt{NaNs} can be propagated through noise-mitigation pipelines with the idea of understanding how many extra quiet \texttt{NaNs} are detected downstream.}.   

For the following discussion, we will define 
\begin{itemize}
\item \textbf{Valid Data:} The contiguous set of defined numerical data that are not corrupted by \texttt{NaNs}.
\item \textbf{Missing Data:} Specific elements of data that are undefined and cannot be represented by floating point numbers. We will treat these numerically with \texttt{NaNs}.
\item \textbf{Gap:} The contiguous set of missing data values between two sets of valid data (a missing packet of data is an element of a gap).
\end{itemize}
To simplify the discussion, we will assume that each of the IFOs have missing data at the same time-stamps. Similarly, for now, we will also assume that we only have one set of contiguous \texttt{NaNs}, implying a single gap in the data stream. 

More formally, let $n = n_{0} \in \{0,\ldots,K-1\}$ be the first instance (and discrete index) of a missing data packet in a contiguous sequence of missing scientific telemetry data (IFO measurements) of length $N^{\text{ifo}}_{\texttt{NaNs}}$. 
This means that all of the IFO measurements will exhibit missing data for indices $n \in \mathcal{S}^{\text{ifo}}_{\texttt{NaNs}}$,
\begin{equation}
\mathcal{S}^{\text{ifo}}_{\texttt{NaNs}} = \{n_0,n_0 + 1,\ldots,n_0 + N^{\text{ifo}}_{\texttt{NaNs}} - 1\}, \label{main_text_eq:telemetry_nans_set}
\end{equation}
such that
\begin{align}\label{main_text_eq:definition_ifo_missing_data}
    \text{ifo}_{ij}[n\in \mathcal{S}^{\text{ifo}}_{\texttt{NaNs}}] &=\text{isi}_{ij}[n\in \mathcal{S}^{\text{ifo}}_{\texttt{NaNs}}] \nonumber\\&= \text{tmi}_{ij}[n\in \mathcal{S}^{\text{ifo}}_{\texttt{NaNs}}]  \\ &= \text{rfi}_{ij}[n\in \mathcal{S}^{\text{ifo}}_{\texttt{NaNs}}] \nonumber \\ &= \texttt{NaNs}. \nonumber
\end{align} 
The string of equalities in Eq.\,\eqref{main_text_eq:definition_ifo_missing_data} should be interpreted as a loss of communication between the instrument and ground-segment. This may not be the case in reality. For instance, we may find that the IFOs on board S/C 1 may exhibit significant glitchy behavior that results in total data loss, whereas the IFOs on S/C 2 and 3 operate at normal level. 
Until more realistic LISA-based simulations are conducted, we cannot derive a complete scenario of  possible collections of missing data (gaps). For this reason, and since this work will form the basis for more detailed missing-data studies on noise mitigation pipelines (pre global-fit), we will assume that all the IFOs measurements are corrupted and lost at the same time-stamps. We will discuss how one could compute the extra missing data due to TDI calculations when considering more general scenarios in later stages of our work. 
\subsection{Gap Augmentation in the Intermediary Variables}\label{subsec:intermediary_variables_gap_augmentation}

The goal here is to understand how many extra missing data values are present in the intermediary variables $\eta_{ij}$ when a fixed number of missing data exists in the interferometric measurements.

Recall from Eq.\,\eqref{main_text_eq:definition_ifo_missing_data} that each of the IFO measurements exhibit the same missing data at the same time stamps. From the perspective of missing data values, this implies that the computed intermediary variables $\eta_{ij}$ will have missing data at the same time-stamps. In other words, defining the set $\mathcal{S}^{\eta}_{\texttt{NaNs}}$ as the ordered set of missing-data values in the intermediary variables, then from Eqs. \eqref{eq:etas_variables} and \eqref{main_text_eq:definition_ifo_missing_data}, we must have for static instrument configurations that $\eta_{ij}\left[n\in \mathcal{S}^{\eta}_{\texttt{NaNs}}\right] = \texttt{NaNs}$ for all six S/C pairings. This means that we only need to focus on computing the gap augmentation in a single intermediary variable, since all other intermediary variables will host the same set of \texttt{NaNs} at precisely the same indices.  

We can re-write Eq.~\eqref{eq:xi_variables} and Eq.~\eqref{eq:etas_variables} to focus on the individual components hampered by missing data. The goal here is to understand the gap augmentation of the following equation. 
\begin{equation}\label{eq:eta_nans_simplified}
\hat{\eta}_{ij}(t) \approx \text{ifo}_{ij}(t) + \delay_{ij}\text{ifo}_{ij}(t)\,.
\end{equation}

We name the standard $\eta_{ij}$ variables as $\hat{\eta}_{ij}$ since we do not consider the specific numerical IFO measurements but the overall number of missing data (\texttt{NaNs}) in our calculations. This will simplify the expressions that feed into the TDI variables. 
In the discrete domain, we need to compute 
\begin{equation}\label{eq:reduced_eta_nans_delay}
\hat{\eta}_{ij}[n] \approx \text{ifo}_{ij}[n] + \sum_{k = -p + 1}^{p}\text{ifo}_{ij}[n - N + k]\,.
\end{equation}
Since we are only interested in the number of extra missing data, we can neglect $\ell_{k}(-\epsilon)$ since it will be unnecessary in our calculations. 

Following Eq.~\eqref{main_text_eq:telemetry_nans_set} and Eq.~\eqref{main_text_eq:definition_ifo_missing_data}, let $N^{\text{ifo}}_{\texttt{NaNs}}$ denote the number of \texttt{NaNs} in the telemetry variables. For both constant and equal arm-lengths, let $\overline{L}$ represent the length of a single LISA link, $c$ the speed of light and $\Delta t$ the sampling interval. The integer delay defined in Eq.~\eqref{eq:reduced_eta_nans_delay} is then given by $N = \lfloor \overline{L}/(c\Delta t)\rfloor$. As derived in Appendix \eqref{app:gap_augmentation_eta}, the overall gap augmentation in the intermediary variables ( Eq. \eqref{eq:reduced_eta_nans_delay}) is given by the piecewise function 
\begin{widetext}
\begin{equation}\label{eq:gap_augmentation_eta_analytical}
N^{\text{ifo} \rightarrow \eta}_{\texttt{NaNs}} = f(\mathcal{K}, N^{\text{ifo}}_{\texttt{NaNs}}, N) = \begin{cases}
    \mathcal{K} + N^{\text{ifo}}_{\texttt{NaNs}} & \text{if} \  1 < \mathcal{K} \leq 1 + 2N - 2N^{\text{ifo}}_{\texttt{NaNs}}\,, \\
    \frac{\mathcal{K} + 1}{2} + N  &1 + 2N - 2N^{\text{ifo}}_{\texttt{NaNs}} \leq \mathcal{K} \leq 2N - 1\,, \\
    \mathcal{K}&  \mathcal{K} \geq 2N -1 \,,
\end{cases}
\end{equation}
\end{widetext}
for $\mathcal{K}$ the order of the Lagrange interpolant for the $\eta_{ij}$ variables. The overall number of missing-data values in the time-series is given by
\begin{equation}\label{main_body_eq:total_number_nans_eta}
N^{\eta}_{\texttt{Nans}} = N^{\text{ifo}}_{\texttt{NaNs}} + N^{\text{ifo} \rightarrow \eta}_{\texttt{NaNs}} \,.
\end{equation}
We remark in general that if one were to use the more numerically stable (and efficient) interpolant found in Ref.~\cite{Staab:2024nyo}, the result given in Eq.\,\eqref{eq:gap_augmentation_eta_analytical} will remain unchanged. The interpolating scheme in Ref.~\cite{Staab:2024nyo} requires far fewer coefficients (22) than the Lagrange interpolants used throughout this work (with default Lagrange interpolation order set to be 45). This would imply that the gap augmentation would be less, since we use less terms to approximate the delayed sample.

We remind the reader that the calculations leading to Eq.\,\eqref{eq:gap_augmentation_eta_analytical} have assumed a static and equal arm-length constellation. In the more general case of time varying arm-lengths, the value of the integer time-shift $N$ will now depend on (1) a specific instance of time and (2) the S/C pair $(ij)$. In other words, $N:=N^0_{ij}[n]$ as reported in Eq.\,\eqref{eq:approx_delay_x}. The LISA arm lengths vary by approximately \(3\%\) for the entire course of the LISA mission~\cite{Martens:2021phh}. As a result, depending on the chosen simulation start time and duration, the mean arm length can take values as large as \(\bar{L}_{\mathrm{max}} = 2.54\times 10^{9}\,\mathrm{m}\) or as small as \(\bar{L}_{\mathrm{min}} = 2.455\times 10^{9}\,\mathrm{m}\). The integer part of the delay $N = \lfloor \bar{L}/c\Delta t\rfloor \in \{32,33\}$ for $\bar{L}\in\{\bar{L}_{\rm min},\bar{L}_{\rm max}\}$ respectively. In contrast, for our numerical simulations in our results section \eqref{subsec:results_verification}, for equal and constant arm-links with $\bar{L} = 2.5\times 10^{9}\,$m, one obtains $N = 33$ for the integer part of the delay. Making reference to Eq.\,\eqref{eq:gap_augmentation_eta_analytical}, the integer time-shift primarily determines which of the three gap augmentation regimes that must be considered when predicting the overall widening of the gap. This means that our assertion that $\bar{L} = 2.5\times 10^{9}\,$m will manifest as an overestimation of the gap widening \emph{by a single data point} per gap at most. This is in the very specific case that we have an order of interpolant $\mathcal{K}$ and original gap length $N^{\rm ifo}_{\texttt{NaNs}}$ that satisfy the intermediary regime set in Eq.\,\eqref{eq:gap_augmentation_eta_analytical}, the regime where the integer time-shift determines the size of the gap widening.

To conclude, we strongly believe that our results (and conclusions) on our scientific sections \ref{sec:Results} will not change if we were to consider more complex orbital configurations.

In the next section, we compute the overall missing data in the first and second generation TDI variables.

\subsection{Gap augmentation in TDI Michelson variables}\label{subsec:gap_augmentation_TDI_variables}

We will assume we have a single set of contiguous missing data in the telemetry variables, resulting in a single gap. We remind the reader since the telemetry variables all have the same missing data at the same time-stamps, we understand that each of the 6 intermediary variables will all have missing data at the same time-stamps. A result of this is that the missing data in the TDI variables $\{\mathbf{X}_1,\mathbf{X}_2\}$ will also be the same as $\{\mathbf{Y}_1,\mathbf{Y}_2\}$ and $\{\mathbf{Z}_1,\mathbf{Z}_2\}$. As a result, we will only provide the calculation for the TDI $\mathbf{X}$ variables, since the other variables for $\textbf{Y}$ and $\textbf{Z}$ follow trivially. 

A computationally favorable scheme for computing both generation of TDI variables $\textbf{X}_{1}$ and $\textbf{X}_{2}$ defined in Eq.~\eqref{eq:1st_gen_TDI_factorised} and Eq.~\eqref{eq:2nd_gen_TDI_factorised} is the following \cite{Staab_2025}. First we can define the quantities
\begin{subequations}
\label{eq:a_variables}
\begin{align}
a_{1} &= \eta_{12} + \delay_{12}\eta_{21} ,\label{eq:a_1} \\
a_{2} &= \eta_{13} + \delay_{13}\eta_{31}\label{eq:a_2},
\end{align}
\end{subequations}
that can be understood as single photon-path round-trip. Then defining second photon-path round-trip
\begin{subequations}
\label{eq:r_variables}
\begin{align}
r_{1} &= a_{1} + \delay_{121}a_{2} ,\label{eq:r_1}\\
r_{2} &= a_{2} + \delay_{131}a_{1}, \label{eq:r_2}
\end{align}
\end{subequations}
it is possible to compute first generation TDI variables 
\begin{align}
\mathbf{X}_{1} &= r_{2} - r_{1}\,. \label{eq:first_gen_factorized_variables} 
\end{align}
Notice that to build $\mathbf{X}_{1}$ we require only 4 interpolation schemes rather than 6 required for the non-factorized expressions first generation TDI expressions \cite{PhysRevD.105.122008,PhysRevD.109.043040}.

The second generation variables require a final set of variables 
\begin{subequations}
\label{eq:q_variables}
\begin{align}
q_{1} & = (1 + \delay_{12131})r_{1} \label{eq:q_1}, \\
q_{2} & = (1 + \delay_{13121})r_{2} \label{eq:q_2},
\end{align}
\end{subequations}
which can be combined to give the second generation TDI variables
\begin{align}\label{eq:second_gen_factorized_variables} 
\mathbf{X}_{2} &= q_{1} - q_{2},
\end{align}
which require only six interpolations, rather than 14 interpolations required to compute \emph{unfactorized} TDI variables. This implies that the cost of computing the second generation factorized TDI variables is significantly less than the unfactorized quantities. We discuss in Appendix \eqref{app:nested_delays_formulas} how to compute the gap augmentation on factorized variables that make use of quantities such as
\begin{equation}\label{main_body_eq:y_nested_delay_x}
y(t) = x(t) + \delay_{i_0i_1\ldots i_n}x(t).
\end{equation}
Notice that each of Eq.\,\eqref{eq:a_variables}, Eq.\,\eqref{eq:r_variables} and Eq.\,\eqref{eq:q_variables} are of a similar form to Eq.~\eqref{main_body_eq:y_nested_delay_x} provided each have missing data at the same time-stamps (as assumed previously). For nested delays, we can use the formula \eqref{eq:gap_augmentation_x_y_nested} defined in Appendix \eqref{app:nested_delays_formulas}, which is a generalization of Eq.\,\eqref{eq:gap_augmentation_eta_analytical} for multiple delay operators. The goal then is to (repeatedly) compute the gap augmentation as Eq.\,\eqref{main_body_eq:y_nested_delay_x} with the gap augmentation given by $$N^{x\rightarrow y}_{\texttt{NaNs}} = f(\mathcal{K},N^{x}_{\texttt{NaNs}},\lfloor n \cdot \bar{L}/(c\Delta t)\rfloor),$$ for $n$ the number of delay operators for each of Eq.\,\eqref{eq:a_variables}, Eq.\,\eqref{eq:r_variables} and Eq.\,\eqref{eq:q_variables} and $N^{x}_{\texttt{NaNs}}$ the number of missing-data present in the $x$ variables is determined by Eq.\,\eqref{main_body_eq:y_nested_delay_x}.

We begin by computing the gap augmentation in each of the $a_{1}$ and $a_{2}$ single round-trip variables defined in Eq.\,\eqref{eq:a_1} and Eq.\,\eqref{eq:a_2}. Since each IFO has missing data in the precise same time-stamps, this implies that the gap augmentation in $a_1$ should be identical to $a_2$. The gap augmentation on both $a_1$ and $a_2$ is given by the following quantity
\begin{equation}\label{eq:gap_augmentation_a1a2_analytical}
N^{\eta \rightarrow a_{(1,2)}} = f(\mathcal{K},N^{\eta}_{\texttt{NaNs}}, \lfloor \bar{L}/(c\Delta t\rfloor)\, ,
\end{equation}
with overall number of missing data in both $a_{1}$ and $a_{2}$, is given respectively by
\begin{equation}\label{eq:gap_total_a1a2_analytical}
N^{a_{(1,2)}}_{\texttt{NaNs}} = N^{\eta}_{\texttt{NaNs}} + N^{\eta\rightarrow a_{(1,2)}}\,,
\end{equation}
with $N^{\eta}_{\texttt{NaNs}}$ given by Eq.\,\eqref{main_body_eq:total_number_nans_eta}. In a similar manner, we can use Eq.\,\eqref{eq:gap_augmentation_x_y_nested} to compute the gap augmentation for $r_1$ and $r_2$. We thus have
\begin{equation}\label{eq:gap_augmentation_r1r2_analytical}
N^{a_{(1,2)} \rightarrow r_{(1,2)}} = f(\mathcal{K},N^{a_{(1,2)}}_{\texttt{NaNs}}, \lfloor 2\bar{L}/(c\Delta t)\rfloor)\,,
\end{equation}
with total number of \texttt{NaNs} present in the $r_1$ and $r_2$ variables 
\begin{equation}\label{eq:gap_total_r1r2_analytical}
N^{r_{(1,2)}}_{\texttt{NaNs}} = N^{a_{(1,2)}}_{\texttt{NaNs}} + N^{a_{(1,2)}\rightarrow r_{(1,2)}}. 
\end{equation} 
 
For second generation variables, we have a final set of nested delays that require computing in Eq.\,\eqref{eq:second_gen_factorized_variables}. Given the overall number of NaNs present in the round-trip variables $N^{r_{(1,2)}}_{\texttt{NaNs}}$, we compute
\begin{align}
N^{r_{(1,2)} \rightarrow q_{(12)}} &= f(\mathcal{K},N^{r_{(1,2)}}_{\texttt{NaNs}}, \lfloor 4\bar{L}/(c\Delta t\rfloor) \label{eq:gap_augmentation_q1q2_analytical}\\
N^{q_{(1,2)}}_{\texttt{NaNs}} &= N^{r_{(1,2)}}_{\texttt{NaNs}} + N^{r_{(1,2)}\rightarrow q_{(1,2)}}\label{eq:total_gap_augmentation_q1q2} \,,
\end{align}
where Eq.\,\eqref{eq:gap_augmentation_q1q2_analytical} determines the overall gap augmentation in the $q_{1}$ and $q_{2}$ variables. By construction, these have the same missing data placements as the second-generation TDI variables $\mathbf{X}_2$.

The quantities Eqs.\eqref{eq:gap_augmentation_eta_analytical}, \eqref{eq:gap_augmentation_a1a2_analytical}, \eqref{eq:gap_augmentation_r1r2_analytical} and finally Eq.\,\eqref{eq:gap_augmentation_q1q2_analytical} can be used to find out the total number of missing data in both first and second generation TDI variables. Combining each of these equations leads to the main result of our paper Eq.\,\eqref{eq:Widening_of_X1_X2_subequations}:
\begin{tcolorbox}[colframe=black, colback=white, arc=4mm, boxsep=0mm, left=2mm, right=1mm, top=0cm, bottom=4mm]
\begin{subequations}\label{eq:Widening_of_X1_X2_subequations}
\begin{align}
N^{\textbf{X}_1}_{\texttt{NaNs}} &= N^{\text{tel}}_{\texttt{NaNs}} + N^{\text{tel}\rightarrow \eta}_{\texttt{NaNs}}  +  N^{\eta \rightarrow a_{(1,2)}}_{\texttt{NaNs}}  + N^{a_{(1,2)} \rightarrow r_{(1,2)}}_{\texttt{NaNs}}, \label{eq:gap_augmentation_first_gen_X} \\[5pt]
N^{\textbf{X}_2}_{\texttt{NaNs}} &= 
N^{\textbf{X}_{1}}_{\texttt{NaNs}} + N^{r_{(1,2)} \rightarrow q_{(1,2)}}_{\texttt{NaNs}} \,. 
\label{eq:gap_augmentation_second_gen_X}
\end{align}
\end{subequations}
\end{tcolorbox}

The two Eqs.(\ref{eq:gap_augmentation_first_gen_X} -- \ref{eq:gap_augmentation_second_gen_X}) determine the overall gap augmentation in both the first- and second-generation TDI variables.

\section{Results}\label{sec:Results}
We begin by verifying our analytical results derived in Sec.\eqref{subsec:gap_augmentation_TDI_variables}, specifically Eqs.\eqref{eq:Widening_of_X1_X2_subequations}, using publicly available LISA simulation codes. Once we show in Sec.\eqref{subsec:results_verification} that our cheap-to-compute gap augmentation codes are faithful to LISA-based simulations, we will upscale our time-series and consider a full four year LISA-like data set with a realistic missing data and gap scenario in Sec.\eqref{subsec:results_full_scale_TDI_simulation_gaps}. We make use of our analytical results since with the current version of LISA instrument we cannot run a four-year long simulation due to memory usage limitations. We will then compute the corrected duty-cycle when accounting for the gap augmentation due to TDI. 
\subsection{Verification}\label{subsec:results_verification}
\begin{figure*}
    \centering
    \includegraphics[width=\linewidth]{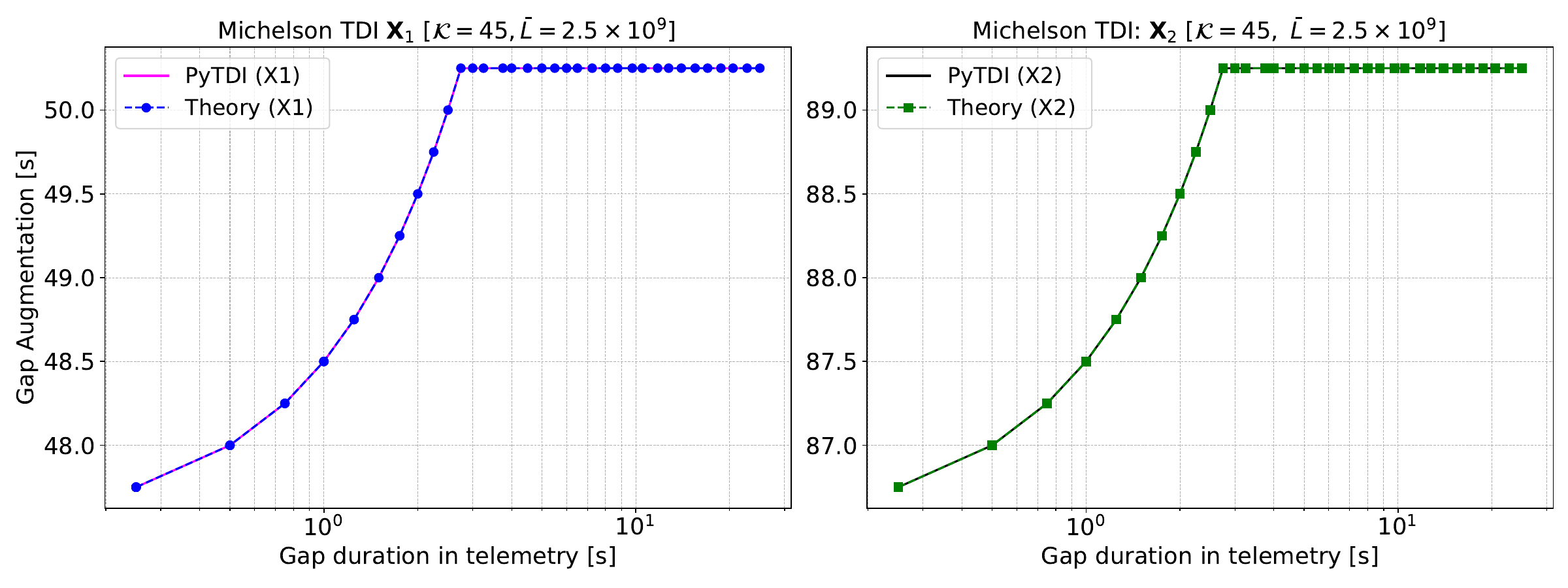}
    \includegraphics[width=\linewidth]{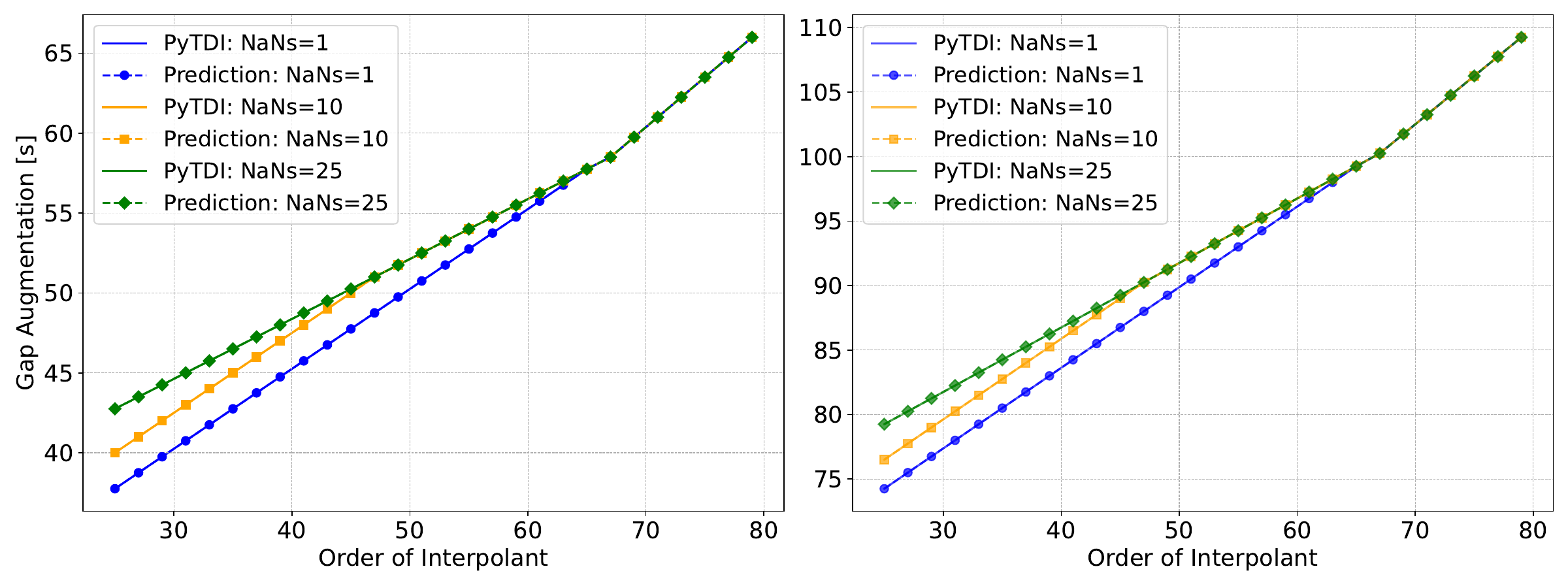}
    \caption{ \textbf{(Top Left/Top right plot):} For first/second generation TDI variables( Eq.\eqref{eq:first_gen_factorized_variables}) we compute the gap augmentation with $\mathcal{K} = 45$ for all Lagrange interpolants as a function of the gap duration. The purple/blue-dashed curves in the top left plot compare the numerical/theoretical gap augmentation in the first generation TDI variables using \texttt{lisainstrument} and \texttt{pyTDI} and Eq.\,\eqref{eq:gap_augmentation_first_gen_X} respectively. The second plot is the same but for second generation TDI variables with black/green curves depicting the gap augmentation using \texttt{pyTDI} and Eq.\,\eqref{eq:gap_augmentation_second_gen_X}. \textbf{(Bottom Left/Bottom right plot):} Here we compare the Gap Augmentation using \texttt{lisainstrument} and \texttt{pyTDI} against our analytical result of Eq.\,\eqref{eq:gap_augmentation_first_gen_X} (left plot) and of Eq.\,\eqref{eq:gap_augmentation_second_gen_X} (right plot) as a function of the order of interpolant (set equal for all delay operators). As we can see, there exist three regimes where the gap augmentation changes slope with respect to the order of interpolant (see Eq.\,\eqref{eq:gap_augmentation_eta_analytical} for similar behaviour with justification outlined in Appendix \ref{app:nested_delays_formulas}). As one can see, our analytical result (dashed blue, yellow and green dots) matches the numerical simulations (solid blue, yelow and green curves) for both plots for various sets of contiguous nans $N^{\text{ifo}}_{\texttt{NaNs}} = \{ 1, 10, 25\}$. A \texttt{jupyter} notebook with code to reproduce these plots is available \href{https://github.com/OllieBurke/gaps_tdi}{here}.}
    \label{fig:TDI_Variables_X_numerical_plot}
\end{figure*}
\subsubsection{Simulation set-up}
In this section, we will make use of \texttt{lisainstrument} to generate a small-scale LISA-based simulation for a static constellation with equal arm-lengths set to $\bar{L} = 2.5 \cdot 10^{9}$\,m for each arm-link. We will consider a sampling rate $f_{s} = 4\,\text{Hz}$ and disable all noise sources except from laser, TM and OMS noise. The overall simulation will be around $8.3\,$ hours in duration. 

From the LISA simulation, we can extract the relevant scientific telemetry data. Specifically, we can obtain the measurements of the six interferometers for each craft. To construct segments of missing data, we will use a binary mask $w[n]$ with $w[n] = 1$ for $n\Delta t \not \in T_{\text{gap}}$ and $w[n] = \texttt{NaN}$ for $n\Delta t \in T_{\text{gap}}$. Here $T_{\text{gap}}$ is the time-interval of the missing data values. We enforce a gap in the telemetry variables by multiplying each of the $\text{ifo}_{ij}$ measurements $(\text{isi}_{ij}, \text{tmi}_{ij} \ \text{and} \  \text{rfi}_{ij})$ by the masking function $w$. We do not apply a mask to the IFO sidebands and MPRs since we do not account for clock noise effects or evolving LISA arm-lengths.

These masked telemetry data are fed into \texttt{pyTDI} to compute the intermediary variables and resultant first and second generation TDI X variables. When computing the intermediary and TDI variables, we fix the order of interpolants to $\mathcal{K} = 45$. For a variable number of missing data (size of binary mask), we compare the result of our analytical calculation Eq.\,\eqref{eq:Widening_of_X1_X2_subequations} with the numerical gap augmentation outputted using PyTDI. 

In our analysis we do not correct for the group delay introduced when the phasemeter data are filtered and decimated from $80\,$MHz to $4\,$Hz. Compensating for this delay would require a time-shift of the telemetry, which in turn would increase the number of missing samples by an amount $\mathcal{A}$, the order of the Lagrange interpolant used for the shift \cite{PhysRevD.99.084023}. These extra gaps would then propagate through \eqref{eq:Widening_of_X1_X2_subequations} and \eqref{eq:gap_augmentation_eta_analytical}. Since our focus here is on gap growth in the scientific data rather than the full noise-reduction pipeline, we ignore this effect; the additional widening is small and does not change our conclusions.

\subsubsection{Discussion}
Our result is given in Fig.~\ref{fig:TDI_Variables_X_numerical_plot}. In all four panels we show that our analytical result of Eq.\,\eqref{eq:Widening_of_X1_X2_subequations} is consistent with the numerical gap augmentation given using both \texttt{lisainstrument} and \texttt{pyTDI}. In the upper plots of the two panels in Fig.~\ref{fig:TDI_Variables_X_numerical_plot}, we report the gap augmentation with respect to the number of \texttt{NaNs} (or equivalently to the duration of the gap) when a fixed interpolation order of 45 has been chosen. Moreover, on the bottom of the two panels
we also analyse three scenarios with 1, 10, and 25 \texttt{NaNs} and we study the gap augmentation with respect to the order of interpolation.

On the top two plots, for both first- and second-generation TDI variables (\(\mathbf{X}_1\) and \(\mathbf{X}_2\)), for gaps larger than 2.5\,s (or equivalently for number of NaNs $>$ 10), the augmentation becomes constant and equal to 
\begin{itemize}
    \item 50.25\,s for first generation variables,
    \item 89.25\,s for second generation variables.
\end{itemize}
This is interesting since it shows that the gap augmentation is not strongly dependent on the gap duration; e.g., for gap durations $\geq 2.5$\,s, the gap augmentation appears to be constant. 

\textbf{Frequent and short gaps may be the most concerning.} 
A loss of a single packet of data ($\sim 0.25$\,s) can lead to an additional loss of $\sim 86.75$\,s in TDI $\mathbf{X}_2$. 
This is particularly relevant for LISA if very short but highly frequent data gaps occur (e.g., antenna-related gaps last about a few hours while PAAM-related gaps are typically of 100\,s), as the cumulative excess loss from TDI could become significant. \textbf{By contrast, less frequent and longer gaps are less affected by TDI.} 
While TDI $\mathbf{X}_2$ can extend a gap by $\sim 90$\,s, this is negligible compared to scheduled long gaps (e.g., estimation of TTL coupling coefficients), which can last up to $\sim 40$ hours. 
Although any additional loss of data is undesirable, its relative impact in this case is minimal. \textbf{In summary,} when comparing losses from TDI operations to those caused by the gaps themselves, short and frequent gaps pose the greatest risk. 
For example, short gaps are expected to occur frequently in the planned schedule (1095 per year, $\sim 100$\,s each), making the TDI-induced loss comparable to the intrinsic loss. 
Conversely, for long gaps, which are rare (6 per year), the additional TDI loss is negligible compared to their planned duration.

We will now conclude this section with a small discussion focused on more generic orbital motions of the craft. As discussed previously, the LISA arm lengths vary by at most $\sim 3\%$, resulting in an integer time-shift (see Eq.\,\eqref{eq:gap_augmentation_eta_analytical}) $N\in\{32,33\}$. The situation becomes a touch more interesting when considering multiple delays, with overall delay expressed by  

\begin{equation}\label{eq:integer_delay_results_discussion}
\left\lfloor \frac{m\,\bar{L}}{c\,\Delta t} \right\rfloor.
\end{equation}
For multiple delays that determine the pre-factor ($n$) in Eq.\,\eqref{eq:integer_delay_results_discussion}, the choice of $\bar{L}\in[\bar{L}_{\rm min}, \bar{L}_{\rm max}]$ will govern smaller/larger values of the integer time-shift. Where here $\bar{L}_{\rm min}$ and $\bar{L}_{\rm max}$ can be estimated from the pre-computed orbits given in Ref \cite{Martens:2021phh}. The value of this integer time-shift determines the regime of gap augmentation located in Eq.~\eqref{eq:gap_augmentation_x_y_nested}. Note that the only regime where the integer time-shift features directly in the gap augmentation is the secondary regime in Eq.\,\eqref{eq:gap_augmentation_x_y_nested}. For second generation TDI variables, we can compute $\lfloor 4 \bar{L}/(c \Delta t)\rfloor \in[131, 135]$ for $\bar{L}\in\{\bar{L}_{\rm min},\bar{L}_{\rm max}\}$, to compare our case, when assuming $\bar{L} = 2.5\times 10^{9}\,$m, this results in $\lfloor 4 \bar{L}/(c \Delta t)\rfloor = 133$. As one can see, the gap augmentation could either increase/decrease (by only a few points) only if our order of interpolant, and number of NaNs present in any of the steps previous to the TDI calculations, force us to fall within the intermediary regime (see for example $r, a,\eta$ in Eqs.\eqref{eq:gap_augmentation_eta_analytical}, \eqref{eq:gap_augmentation_a1a2_analytical} and \eqref{eq:gap_augmentation_r1r2_analytical}).

We have indeed checked our analytical formula for more generic orbits of LISA using \texttt{lisainstrument} and \texttt{pyTDI} with $\bar{L} \in [L_{\rm min}, L_{\rm max}]$ and our observations were expected. For $\mathbf{X}_{1}$ and $\mathbf{X}_{2}$, we found that we under-estimated the overall gap augmentation by around one and three points respectively, if $\bar{L} = \bar{L}_{\rm max}$ or $\bar{L} = \bar{L}_{\rm min}$ for a short $\sim 8.3\,\text{hour}$ simulation. For longer LISA-based simulations, the arm-lengths are slowly oscillating around a mean-value so the gap augmentation can decrease as well as increase. We believe that our average value is a solid baseline to make scientific conclusions out of. As a result, we strongly believe that our conclusions would not change significantly if we were to consider more general orbits of the LISA constellation.     

\subsection{LISA-based Gap Simulations}\label{subsec:results_full_gaps_telemetry}

 It is expected that LISA should operate normally for four years with a quoted $\sim 82\%$ of usable data, the so-called \emph{duty cycle} of the mission \cite{colpi2024lisadefinitionstudyreport}. Generating a full four-year long LISA simulation sampled at 4\,Hz using both \texttt{lisainstrument} and \texttt{pyTDI} is not only extremely expensive but prohibitively memory intensive. Instead, using the results of the previous section and the details outlined in Appendix \eqref{app:multiple_gaps_telemetry_variables}, one can compute the overall extension of missing data in TDI variables relatively cheaply. Our results can inform performance and operation personnel on the loss of data in case gaps are present in the raw phasemeter data and TDI-augmented.   

Tab.\eqref{tab:gap_summary}, inspired by Ref.\cite{Castelli:2024sdb} and extended to include additional gap scenarios arising from the design of LISA, outlines a potential situation in which the spacecraft is affected by both \textit{planned} and \textit{unplanned} data gaps. Here, we define \textit{planned} gaps as those due to routine instrumental maintenance, whereas \textit{unplanned} gaps correspond to unexpected instrumental malfunctions or complete communication losses caused by environmental effects, leading to a total loss of usable data (i.e., not attributable to routine maintenance). 

Such malfunctions can either produce excessive noise bursts that are subsequently removed from the data (glitch subtraction) or cause communication outages leading to the loss of interferometric variables required for TDI computations. These artifacts may result in corrupted data that cannot be used in parameter inference pipelines. While such data segments can be entirely removed (masked), this approach enables the application of simpler and more cost-effective search and parameter estimation techniques, as demonstrated in \cite{Burke:2025bun,Dey:2021dem,Castelli:2024sdb}.

    \begin{table*}
    \centering
    \renewcommand{\arraystretch}{1.1}
    \setlength{\tabcolsep}{8pt}
    \begin{tabular}{l l c c c c c}
        \hline\hline
        Gap Type & Subtype & Rate [yr$^{-1}$] & Duration [hr] & (masked) Segments & Total Hours & Duty Cycle [\%] \\
        \hline
        \multirow{3}{*}{Planned} 
          & Long & 6 & 40 & \multirow{3}{*}{4339} & \multirow{3}{*}{1363.38} & \multirow{3}{*}{96.112} \\
          & Medium & 26 & 3.3 & & & \\
          & Short & 1095 & 0.028 & & & \\
        \hline
        \multirow{3}{*}{Unplanned} 
          & Platform & 3 & 60 & \multirow{3}{*}{126} & \multirow{3}{*}{4156.82} & \multirow{3}{*}{88.145} \\
          & Payload & 4 & 66 & & & \\
          & Environmental & 30 & 24 & & & \\
        \hline
        \makecell[c]{Planned \\ + \\ Unplanned} & \qquad \ \ — & — & — & 3944 & 5367.04 & 84.257 \\
        \hline\hline
    \end{tabular}
    \caption{The first column represents the class of gaps, be them Planned, Unplanned or a mixture of both. The second column represents the subtype of gap with third and fourth columns giving expected rates/year and durations [hours]. The fifth, sixth and seventh columns represent the median number of gap segments, total hours of missing data and overall duty cycle for $\sim 2000$ seeded masking simulations sampled at $\Delta t = 10\,$s with assumed four-year mission duration. The median total duty cycle reflects the proportion of the year uncontaminated by gaps. The distributions of the duty cycles are given in Fig.~\ref{fig:duty_cycle_distribution}. We note that the total number of gated segments in the planned+unplanned configuration is smaller than in the planned-only case, since some of the more frequent but shorter planned data gaps are absorbed into the longer and less frequent unplanned gaps.}
    \label{tab:gap_summary}
\end{table*}

We sample the \textit{planned} data gaps at regular intervals following the Rate [yr$^{-1}$] given  in Table (\ref{tab:gap_summary}), with a $\sim10\%$ random jitter applied to (the start time) the gap duration of each subtype of gap. For the \textit{unplanned} data gaps, the gap duration (start times) are sampled from an exponential distribution which is derived again accordingly to the Rate [yr$^{-1}$] given in Table (\ref{tab:gap_summary}). The result is a binary mask with integer valued $1$ representing valid data and \texttt{NaNs} representing missing data. 

In the following, we will assume that the instrument loses data in all IFO measurements at precisely the same time-stamps. We will construct three masks, one for either planned, unplanned and with both planned and unplanned with rate parameters and durations given by the second and third column of Tab.~(\ref{tab:gap_summary}). 

\begin{figure*}
    \includegraphics[width = \textwidth]{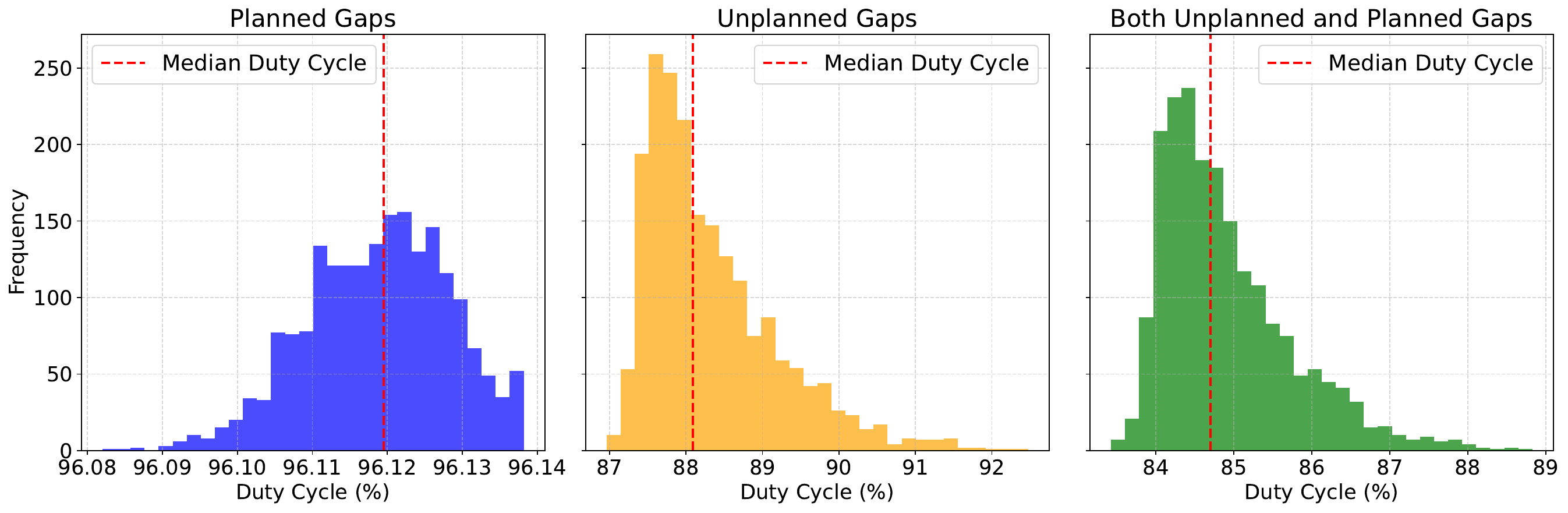}
    \caption{The left/center/right plots show distributions on the resultant duty cycles on the telemetry data when planned/unplanned/both planned and unplanned gap strategies are used via the rates in Tab.~\eqref{tab:gap_summary}.    \label{fig:duty_cycle_distribution}}
\end{figure*}

Based on approximately 2000 seeded realizations of the four-year masking function, sampled at $\Delta t = 10$s with the parameters specified in Tab.\eqref{tab:gap_summary}; Tab.\eqref{tab:gap_summary} summarizes the median number of masked segments, total hours lost, and duty cycles. The distributions on the telemetry duty cycles is given in Fig.(\ref{fig:duty_cycle_distribution}).

What we learn from Tab.~\eqref{tab:gap_summary} and Fig.~\ref{fig:duty_cycle_distribution} is that:
\begin{itemize}
    \item Unplanned data gaps account for the majority loss of data for the instrument, accounting for median $\sim 11.855$\% loss, where this percentage is computed considering 88.145\% of duty cycle,
    \item Planned data gaps only account for a median $\sim 3.888$\%, where this percentage is again computed considering 96.112\% of duty cycle,
    \item Therefore, the overall (median) duty cyle is $100\% - (11.855 + 3.888)\% = 84.257\%$; which is bigger than the quoted $82\%$ nominal duty cycle \cite{colpi2024lisadefinitionstudyreport,Castelli:2024sdb}.
\end{itemize}
This latter point means that we retain approximately 2.257\% $\approx 35.61$ days of worth of scientific data compared to the previously estimated (nominal) 82\% duty cycle \cite{colpi2024lisadefinitionstudyreport,Castelli:2024sdb}. \textbf{The nominal 82\% duty cycle only accounts for the individual gaps affecting the LISA data rather than their combination}\footnote{Note that short duration, high rate gaps may be contained in long-duration, low rate gaps as extensively demonstrated in Appendix \eqref{app:gap_augmentation_eta} and Appendix \eqref{app:nested_delays_formulas}.}, which is outlined in Tab.~\eqref{tab:gap_summary}. This would result in an \textbf{increase of $\sim 35.61$ days to garner more science from GW observations.}
\subsection{Full scale gaps in second generation TDI Michelson variables}\label{subsec:results_full_scale_TDI_simulation_gaps}

We can go further and compute the extra loss of data due to TDI data streams with missing data. We begin by simulating 600 seeded data gaps with rate parameters defined in Tab.~\eqref{tab:gap_summary}, at a sample rate of 4\,Hz. To compute the overall number of missing data on telemetry variables, with arbitrary numbers of gaps and of arbitrary duration, we refer the reader to Appendix \eqref{app:multiple_gaps_telemetry_variables}. Specifically, we will use the equations that lead up to Eq.\,\eqref{app_eq:list_missing_data_sets_X1_X2} to compute the widening of the individual gaps for second generation Michelson TDI variables. Our median results are tabulated in Tab.~\eqref{tab:duty_cycle_widening_full}. 
\subsubsection{Unplanned Gaps}
In this paragraph we will see that although the unplanned data gaps are comparatively much larger than the planned data gaps (see Table \eqref{tab:gap_summary}), the impact of the extra missing data governed through TDI is far less severe than the planned gaps. Results are summarized in Table~\eqref{tab:duty_cycle_widening_full}. The first column reports the type of gap considered, the second column gives the duty cycle of the telemetry data, the third column shows the duty cycle after applying TDI — where some data are lost due to gap augmentation — and the last column presents the difference between the two duty cycles, i.e., the total data loss over the 4-year observation period.

For unplanned data gaps we lose out on $\sim 3$ hours worth of data after applying TDI (see Tab.~\eqref{tab:duty_cycle_widening_full}), a small fraction when compared to the overall loss of data due to the unplanned data gaps. Since the order of interpolant has been fixed to $\mathcal{K} = 45$, the gap widening becomes constant for gaps longer than 2.5\,s. This has been previously depicted in Fig.~\ref{fig:TDI_Variables_X_numerical_plot}, which shows that the gap augmentation becomes constant for gap durations longer than seconds. 

Although the gaps are large comparative to the planned data gaps in duration, the gap widening is constant and small, resulting in an augmentation (after TDI has been applied) of $\sim 90$\,s per gap.

Given the family of unplanned gaps tabulated in Tab.~\eqref{tab:gap_summary},  when not accounting for the merging of the gaps for the unplanned data set, we would expect approximately $4\cdot (30 + 4 + 3)
\cdot 89.25\,\text{s} \sim 0.15$ days lost of data where we have considered the total Rates, the gap augmentation of 89.25\,s for TDI $\mathbf{X}_{2}$ and 4 years of observation.
This is a reasonable (but overestimated) approximation when compared to Tab.\eqref{tab:duty_cycle_widening_full} that shows the true gap augmentation in the TDI variables that accounts for potential merging of gaps.
\subsubsection{Planned Gaps}
For the planned data gaps we see that after applying TDI the overall loss of data is much larger than the unplanned ones, approximately $\sim 4.4852$ days worth, which is approximately $\sim 35$ times more data lost than the unplanned data gaps. This is primarily due to the high-rate and short-duration data gaps with rate $1095/\text{yr}$. Indeed, from a back-of-the-envelope calculation we see that there should be $4 \cdot (6 + 26 + 1095) \cdot 89.25\,\text{s} \approx 4.66$ days worth of lost data. We see from Tab.~\eqref{tab:duty_cycle_widening_full} that our estimate is reasonable, with the true result for the total lost LISA data for the planned scheme being $\approx 4.48$ days, where here we have accounted for gap merging.\\
Finally, the combination of both planned and unplanned data gaps results in a loss of duty cycle of approximately $0.28\%$, which is $\sim$ 4 days worth of data. 

\textbf{To conclude, what we learn here is that gaps of short duration but high rates of occurrence will result in the greatest amount of extra lost data when ran through TDI-based noise-mitigation pipelines. Our results here support the literature that short and frequent gaps may result in a greater impact on the science for LISA than longer duration, but lesser frequent gaps.}  

\begin{table*}[ht]
    \centering
    \renewcommand{\arraystretch}{1.1}
    \setlength{\tabcolsep}{10pt}
    \begin{tabular}{l c c c}
        \hline\hline
        Gap Type & Duty Cycle (Telemetry)[\%] & Duty Cycle (TDI) [\%] & Data Lost [days] \\
        \hline
        Planned & 96.12 & 95.81 & 4.4852 \\
        Unplanned & 88.02 & 88.01 & 0.1291\\
        Planned + Unplanned & 84.65 & 84.37 & 4.0725 \\
        \hline\hline
    \end{tabular}
    \caption{The first column represents the Gap type with gap parameters defined in Tab.\eqref{tab:gap_summary}. The second column is the median duty cycle in the masking function used to mask the telemetry variables. The third column is the overall duty cycle in the TDI variables after the gap augmentation has been taken into account. The final column is the potential loss of science data due to propagating missing-data from the scientific telemetry to the final TDI variables. In this simulation, we use 600 seeded telemetry-based masking functions sampled at 4\,Hz with assuming a mission duration of 4 years.}
    \label{tab:duty_cycle_widening_full}
\end{table*}

\section{Conclusions and future work}\label{eq:conclusions_future_work}

The primary focus of this analysis is to understand the impact that missing raw phasemeter (scientific telemetry) data has on the resultant TDI observations. By treating missing instrumental telemetry data (and sets of missing data -- gaps) in the form of unquantifiable floating points (\texttt{NaNs}), we have shown that ignorantly masking the telemetric data will result in a gap augmentation in the resultant first/second generation TDI streams for $X, Y, Z$. Since the delay operators are computed using FIR (Finite Impulse Response filter) filters, we found that the gap augmentation linearly depends on the order of the kernel (and thus the order of the Lagrange interpolant used in this work). Analytical results were obtained (see Eq.\,\eqref{eq:Widening_of_X1_X2_subequations}) that can cheaply compute the gap augmentation in the intermediary and first/second generation TDI Michelson variables ($\mathbf{X}_1$ and $\mathbf{X}_2$) for any family of gaps with arbitrary durations and rates. For first and second generation TDI variables, with a fixed order of 45 Lagrange interpolant and gap duration $\geq 2.5$\,s, we observe losses of data of duration $\sim 90$\,s and $\sim 50$\,s respectively. The loss of data is greatest for second generation TDI variables, rather than first generation TDI variables, and this is due to the overall number of fractional delay computations (and thus interpolants) applied to the data (with 6 in total applied to second generation rather than 4 for first generation). In Fig.~\ref{fig:TDI_Variables_X_numerical_plot}, we found that the gap augmentation for first and second generation variables is largely insensitive to the gap duration for data gaps of duration longer than 2.5\,s. As demonstrated in Fig.~\ref{fig:TDI_Variables_X_numerical_plot}, we found that a single packet of lost telemetry data results in a loss of TDI-based data lasting approximately $\sim 86.25$\,s. This highlights the severity of short but frequent gaps in the raw phasemeter data, which can lead to substantial data loss once propagated through TDI.  

With analytical results in hand that can cheaply and accurately predict the gap augmentation in TDI variables, we then investigated a LISA-based simulation study on data gaps assuming a realistic gap scenario given by Tab.~(\ref{tab:gap_summary}). \textit{We demonstrated that the median effective duty cycle is $84.6\%$ when accounting for both planned and unplanned data gaps, which is approximately $\sim 2.6\%$ higher than what has been quoted in the literature~\cite{LISA:2024hlh}}. This is due to the realistic case of overlapping gaps in both the planned and unplanned configurations. Averaging over multiple realisations of four year long binary masks sampled at 4\,Hz, we used our cheap-to-evaluate analytical results to quote median gap augmentation results in the second generation TDI variables. Our results were given in Tab.~(\ref{tab:duty_cycle_widening_full}). We concluded that for planned gaps short duration but high rate data gaps lead to the greatest loss of data, resulting in an effective duty cycle loss of $\sim 0.3\%$, which is approximately $\sim 4$ days worth of lost data. 

This loss of data due to TDI is not as severe for the unplanned data gaps (gaps with smaller rates but greater durations), with an overall loss of $\sim 0.01\%$, i.e., $3$ hours worth of lost data. 

High rate and short duration gaps that may appear harmless upstream in the noise mitigation pipeline may have drastic consequences downstream during the global fit analyses. Not only is the loss of scientific data never warranted, but these gaps could also have serious consequences on the search and parameter estimation schemes. Parameter estimation schemes rely on a solid understanding of the statistical properties of the noise process. The presence of many short-duration and high-rate gaps may start to impact assumptions on the process -- in particular whether the between gap segments are independent, dependent. Spectral leakage, if not fully accounted for in the noise covariance matrix could also significantly increase the uncertainty in the parameter posterior~\cite{Burke:2025bun, Edy:2021par, Talbot:2021igi, Baghi:2019eqo}. No matter, what domain, be it time, frequency or time-frequency~\cite{Cornish:2020odn, pearson2025handlingdatagapsgeneration}, all domains of analyses would be negatively impacted by these short duration gaps.

\textit{A clear take home here is that shorter and more frequent gaps will return a greater loss of data than longer but less frequent gaps (with the same duty cycle as the former). An outcome of this research is that, if possible, one should maintain the instrument in such a way to reduce the number of frequent gaps that could occur.} 

There are a number of future projects that could stem from the details contained in this work. For instance, we study here only the current baseline Michelson TDI X. However, other TDI combinations, such as the Sagnac type \cite{Muratore:2020mdf,PhysRevD.105.023009,Hartwig_2022}, might behave differently because they involve shorter delays and therefore gaps are less spread. Similarly, techniques like TDI-Infinity could be better suited to avoid losses due to gaps \cite{houba2024robustbayesianinferencegapped}. Moreover, the top right panel of Figure~(\ref{fig:TDI_Variables_X_numerical_plot}) demonstrates for kernel orders $\mathcal{K} = 45$ that if a single packet of missing data is lost (of duration $\sim 0.25$\,s), then one could lose an significant amount of data in comparison (say, $\sim 86.25$\,s worth). Here we provide a worse case scenario where a single missing data point results in a major loss of data. However, one could use more sophisticated interpolation techniques that are able to account for \emph{unevenly} sampled data points. Smaller sets of contiguous missing-data could be interpolated, or potentially imputed using probabilistic techniques. Interpolating the missing data will force the time-series to be evenly sampled, facilitating the use of spectral methods for downstream signal extraction analyses. \\
 
However, it is of paramount importance that the key idea of TDI is maintained: the overall subtraction and mitigation of laser noise that enables gravitational-wave inference. In practice, this means that any interpolation (or imputation) method is used to recover missing samples must be accurate enough to preserve the laser-noise cancellation. For example, if a single data point is missing and we replace it using interpolation, the reconstructed value must be consistent with what would have been measured so that the subsequent TDI combinations still achieve near-complete laser-noise removal. At the same time, interpolation is performed not only over noise but also over the underlying gravitational-wave signal. If the interpolation is inaccurate, it can distort the signal locally and introduce systematic biases in the recovered waveforms and astrophysical parameters. Therefore, interpolation schemes must simultaneously ensure robust laser-noise cancellation and safeguard the fidelity of gravitational-wave inference.
 
In this work, we have assumed that all telemetry losses occur at the same time-stamps across the interferometers, i.e., all observations are lost simultaneously. In reality, however, data gaps may occur at different times in different interferometer measurements. A natural extension of this study would be to explore the impact of such asynchronous data losses on TDI computations. In this case, the action of delay operators on the intermediary variables would cause each individual gap to widen differently in each interferometer stream. As these propagate downstream, the resulting TDI observables would exhibit multiple gaps, distributed across the $X, Y, Z$ combinations at different time intervals. We expect that this scenario would lead to a greater effective loss of data in the TDI outputs, since the gap segments in different channels are no longer fully overlapping and may become mutually exclusive. From an implementation perspective, this effect could be represented as three distinct binary masks—one for each TDI variable—each containing gap segments at different time positions and with varying durations. This could be easily explored using the LISA simulation tools \texttt{lisainstrument} and \texttt{pyTDI}. We purposefully chose not to explore this further without a clearer understanding of the LISA instrument and what could result in such losses of information. 

Lastly, the impact of clock noise on TDI processing -- which is closely tied to the specifics of the data-processing pipeline-- and therefore on gaps propagation has not been taken into account here. In the case of long-term clock drifts, the relative frequency offsets between spacecraft clocks can accumulate to significant values over the mission lifetime. For example, a relative offset of order $10^{-6}$ leads to a timing error of $\mathcal{O}(100)\,\mathrm{s}$ over four years ($\sim 10^{8}\,\mathrm{s}$). While such drifts appear as common-mode terms in the applied delays and time-stamp offsets and therefore do not strictly need to be corrected prior to TDI, they would result in delays well outside the nominal arm-length scale $\bar{L}$ adopted throughout this work. It is thus preferable to correct for these large drifts, ensuring that the delays remain within a physically meaningful and constrained range. The in-band impact of clock correction depends on the chosen implementation strategy. Two main approaches are considered in the literature: (i) applying the correction jointly with the primary delays and Doppler shifts when constructing laser-noise-reducing TDI combinations~\cite{PhysRevD.105.122008}, or (ii) applying it in a subsequent step after laser-noise reduction~\cite{PhysRevD.103.123027}. In either case, the required clock measurements are typically reconstructed through sensor fusion of sidebands, carrier tones, and possibly additional references such as pseudo-random noise codes, ground-based signals, or TDI-Ranging~\cite{Bayle:2025qvw,PhysRevD.109.022004}. The propagation of telemetry gaps through this reconstruction is non-trivial. For instance, in a simple sensor-fusion scheme, the sideband frequency fluctuations are integrated to recover the clock signal; if a data gap interrupts this process this force the integration to restart.

This work forms the foundations of missing-data on space-based noise-mitigation pipelines, and so we leave these concepts as future work. 

\section{Data Release}
The authors feel strongly about providing open-source software that others can use to build upon this work. The work here was not possible without the use of open sources codes -- in particular \texttt{lisainstrument},  \texttt{pyTDI} and \texttt{lisaglitch}~\cite{bayle2023lisainstrument,Staab2022PyTDI, bayle_2022_6452904}. We release our codes in a Github repository found at  \href{https://github.com/OllieBurke/gaps_tdi}{this link}. 
\section{Acknowledgements}
O.Burke is deeply thankful for Eleanora Castelli for both the wonderful and insightful conversations regarding LISA simulation and for giving O. Burke access to her gap masking code. O. Burke also appreciates the many conversations he had with L. Speri and C. Chapman-Bird (who read and gave excellent comments on an early draft), particularly on gaining perspective on this work. M.Muratore is very grateful for the support given by Olaf Hartwig towards providing a better understanding of the results obtained in this manuscript and to P. Wolf and A. Hees for providing insight on the effect of having gaps when clock noise is taken into account. M.Muratore also acknowledges previous fruitful discussions with A. Hees on gap propagation in TDI during the preparation of the document for the LISA Mission Adoption Review. Last, but certainty not least, M.Muratore acknowledge Wolfrang Kastau for his hint on using NaNs to propagate and keep track of missing data. O.Burke and M.Muratore are deeply thankful for many insightful discussions with the CU L01 group. In particular, O. Burke and M. Muratore are very grateful for the support given by Jean Baptiste-Bayle and Martin Staab. M.Muratore. gratefully acknowledges the support of the German Space Agency, DLR. The work is supported by the Federal Ministry for Economic Affairs and Climate Action based on a resolution of the German Bundestag (Project Ref. No. FKZ 50 OQ 2301). O. Burke and G. Woan acknowledge financial support from the Grant UKRI972 awarded via the UK Space Agency. This work has made use of \texttt{pyTDI}, \texttt{lisaintrument} and \texttt{lisaglitch}.

\appendix 

\section{Derivation of Gap Augmentation for Intermediary Variables}\label{app:gap_augmentation_eta}
\subsection*{Preliminaries}

In our prescription, missing data will be represented via symbolic \texttt{NaNs}, where any binary operation (addition, multiplication, division, subtraction) involving \texttt{NaNs} will return \texttt{NaNs}. Let $n = n_{0}$ be the first instance (and discrete index) of a missing data packet in a contiguous sequence of missing data of length $N^{\text{ifo}}_{\texttt{NaNs}}$. 

This means that all of the IFO measurements will exhibit missing data for indices $n \in \mathcal{S}^{\text{ifo}}_{\texttt{NaNs}}$ with ordered set
\begin{equation}
\mathcal{S}^{\text{ifo}}_{\texttt{NaNs}} = \{n_0,n_0 + 1,\ldots,n_0 + N^{\text{ifo}}_{\texttt{NaNs}} - 1\} \label{eq:telemetry_nans_set}
\end{equation}
 with
\begin{align}
    \text{ifo}_{ij}[n\in \mathcal{S}^{\text{ifo}}_{\texttt{NaNs}}] &=\text{isi}_{ij}[n\in \mathcal{S}^{\text{ifo}}_{\texttt{NaNs}}] \nonumber\\&= \text{tmi}_{ij}[n\in \mathcal{S}^{\text{ifo}}_{\texttt{NaNs}}] \label{eq:definition_ifo_missing_data} \\ &= \text{rfi}_{ij}[n\in \mathcal{S}^{\text{ifo}}_{\texttt{NaNs}}] \nonumber \\ &= \texttt{NaNs}. \nonumber
\end{align} 
We will define 
\begin{align}
n_{\text{first}}^{\text{ifo}} &= n_0 \\
 n_{\text{last}}^{\text{ifo}} &= n_0 + N^{\text{ifo}}_{\texttt{NaNs}} - 1
\end{align}

as the first and last indices of the gapped segment in the telemetry variables. The goal now is to count the missing-data extension (gap augmentation) of the intermediary variables defined in Eq.\,\eqref{eq:reduced_eta_nans_delay}. We are not interested, quantitatively speaking, on the specific values of valid data in the intermediary variables quantitavely speaking. For this reason, we can neglect terms such as $\ell_{k}(-\epsilon)$ in Eq.\,\eqref{eq:approx_delay_x}. Given that
\begin{equation}\label{eq:delayed_eta_using_y}
(\delay_{ij}\text{ifo}_{ij})[n] \approx \sum_{k = -p + 1}^{p}\text{ifo}_{ij}[n - N + k]\,,
\end{equation}
then  the first and final indices (ordered left to right) of the \texttt{NaN} placement in Eq.\,\eqref{eq:delayed_eta_using_y} are given respectively by
\begin{subequations}\label{app_eq:delayed_indices_telemetry}
\begin{align}
n_{\text{first}}^{\delay_{ij}(\text{ifo})} &= n_0 + N - p \\
n_{\text{last}}^{\delay_{ij}(\text{ifo})} &= n_0 + (N^{\text{ifo}}_{\texttt{NaNs}} - 1) + N - (1-p)
\end{align}
\end{subequations}
We can thus write the set of \texttt{NaN} index placements in Eq.\,\eqref{eq:delayed_eta_using_y} 
\begin{align}
\mathcal{S}^{{\delay (\text{ifo})}}_{\texttt{NaNs}} &= \left\{n_{\text{first}}^{\delay_{ij}(\text{ifo})}, n_{\text{first}}^{\delay_{ij}(\text{ifo})} + 1,  \ldots, n^{\delay_{ij}(\text{ifo})}_{\text{last}}\right\}\,. \label{eq:delayed_nans}
\end{align}
Notice that the number of contiguous \texttt{NaNs} between the first and last indices (via an inclusive difference) of the equation above is 
\begin{equation}\label{eq:number_nans_delayed_set}
    \left(n^{\delay_{ij}(\text{ifo})}_{\text{last}} - n^{\delay_{ij}(\text{ifo})}_{\text{first}} + 1 \right) = \mathcal{K} + N^{\text{ifo}}_{\texttt{NaNs}}. 
\end{equation}
The goal now is, given a number of contiguous \texttt{NaNs} in the set $\mathcal{S}^{\text{ifo}}_{\texttt{NaNs}}$ defined in Eq.\,\eqref{eq:telemetry_nans_set}, to compute the gap augmentation in the expression
\begin{equation}\label{eq:reduced_eta_appendix}
\hat{\eta}_{ij}[n] = \text{ifo}_{ij}[n] + \delay_{ij}\text{ifo}_{ij}[n]\,.
\end{equation}
The overall extension of the telemetry missing data (gap augmentation) depends on the overall missing data placement in both the sets $\mathcal{S}^{\text{ifo}}_{\texttt{NaNs}}$. As we will see in the next few subsections, there are three regimes that must be considered that depend on the order of interpolant $\mathcal{K}$, integer time shift $N$ and number of contiguous missing data in telemetry $N^{\text{ifo}}_{\texttt{NaNs}}$. 

\subsection*{Regime 1: No Merging}

Let us suppose that the \texttt{NaN} placements in the telemetry variables do not have the same time-stamps as the delayed telemetry variables defined in \eqref{eq:delayed_eta_using_y}. Mathematically, this can be written using the set notation defined in the previous section 
\begin{subequations}\label{eq:criteria_regime_1}
\begin{align}
&\quad \mathcal{S}^{\text{ifo}}_{\texttt{NaNs}} \cap \mathcal{S}^{\delay_{ij}(\text{ifo})}_{\texttt{NaNs}} = \emptyset  \label{eq:empty_set_nans}\,.
\end{align}
\end{subequations}
In other words, the two sets defined by Eq.\,\eqref{eq:telemetry_nans_set} and \eqref{eq:delayed_nans} are \emph{mutually exclusive} and have no common elements. This means that the extra missing data induced via the delay operator is given by the size of the set \eqref{eq:delayed_nans}. As shown in \eqref{eq:number_nans_delayed_set}, the gap augmentation $N^{\text{tel}\rightarrow\eta}_{\texttt{NaNs}}$ is given by
\begin{equation}
N^{\text{tel}\rightarrow\eta}_{\texttt{NaNs}} = \mathcal{K} + N^{\text{ifo}}_{\texttt{NaNs}}\,. \label{eq:gap_augmentation_eta_regime_1}
\end{equation}
With total number of missing data in the intermediary $\hat{\eta}_{ij}[n]$ variables given by 
\begin{align}
N^{\eta}_{\texttt{NaNs}} &=  \underbrace{N^{\text{ifo}}_{\texttt{Nans}}}_{\text{Telemetry}} + \underbrace{\mathcal{K} + N^{\text{ifo}}_{\texttt{NaNs}}}_{\text{Gap Augmentation}} \nonumber \\ &= \mathcal{K}+ 2N^{\text{ifo}}_{\texttt{Nans}}\label{eq:overall_nans_eta_regime_1} \,.
\end{align}
The gap augmentation \eqref{eq:gap_augmentation_eta_regime_1} and total number of nans \eqref{eq:overall_nans_eta_regime_1} are correct only if the criteria defined in \eqref{eq:criteria_regime_1} are satisfied. The moment that $\mathcal{S}^{\text{ifo}}_{\texttt{NaNs}}\subseteq \mathcal{S}^{\delay_{ij}(\text{ifo})}_{\texttt{NaNs}}$ and  $\mathcal{S}^{\text{ifo}}_{\texttt{NaNs}}\cap \mathcal{S}^{\delay_{ij}(\text{ifo})}_{\texttt{NaNs}} \neq \emptyset$, the overall gap augmentation changes. In the next section, we discuss this in more detail. 

\subsection*{Regime 2: Partial Merging}
We are interested in finding the moment where the set defined in Eq.\,\eqref{eq:delayed_nans} starts to merge with the original telemetry \texttt{NaNs} identified via Eq.\,\eqref{eq:telemetry_nans_set}. This occurs when the last index of the set $\mathcal{S}^{\text{ifo}}_{\texttt{NaNs}}$ equals the first index of the set $\mathcal{S}^{\delay_{ij}(\text{ifo})}_{\texttt{NaNs}}$, i.e., 
\begin{equation}
    S^{\text{ifo}}_{\texttt{NaNs}}\cap S^{\delay_{ij}(\text{ifo})}_{\texttt{NaNs}} =\{n^{\text{ifo}}_{\text{last}} \}=  \{n^{\delay_{ij}(\text{ifo})}_{\text{first}}
    \}\,.
\end{equation}
The choice of Lagrange interpolant allowing this condition is 
\begin{equation}\label{eq:first_transition_point}
\mathcal{K} = 1 + 2N - 2N^{\text{ifo}}_{\texttt{NaNs}}\,.
\end{equation}
Notice the transition point now depends on the choice of integer delay $N$ and number of missing data samples in the telemetry variables $N^{\text{ifo}}_{\texttt{NaNs}}$. Similarly, another transition point is the first instance where the telemetry missing data set is entirely contained within the delayed missing data set. Mathematically, 
\begin{equation}
\mathcal{S}^{\text{ifo}}_{\texttt{NaNs}} \cap \mathcal{S}^{\delay_{ij}(\text{ifo})}_{\texttt{NaNs}} = \mathcal{S}^{\text{ifo}}_{\texttt{NaNs}},
\end{equation}
which will only happen when first telemetry \texttt{NaN} index of $\mathcal{S}^{\text{ifo}}_{\texttt{NaNs}}$ is equal to the first \texttt{NaN} index of $\mathcal{S}^{\delay_{ij}(\text{ifo})}_{\texttt{NaNs}}$, i.e., when $\{n^{\delay_{ij}(\text{ifo})}_{\text{first}}\} = \{n^{\text{ifo}}_{\text{first}}\}$, giving the solution
\begin{align}\label{eq:second_transition_point}
\mathcal{K} = 2N - 1\,.  \quad
\end{align}

The gap augmentation with respect to the telemetry variables is then given by the (inclusive) difference between the last element in $\mathcal{S}^{\delay_{ij}(\text{ifo})}_{\texttt{NaNs}}$ and last element in $\mathcal{S}^{\text{ifo}}_{\texttt{NaNs}}$. A short calculation shows 
\begin{align}
N^{\text{tel}\rightarrow \eta}_{\texttt{NaNs}} = n^{\delay_{ij}(\text{ifo})}_{\text{last}} -  n^{\text{ifo}}_{\text{last}} + 1 = \frac{\mathcal{K} + 1}{2} + N \,.
\end{align}
which is valid under the condition that $\mathcal{K} \geq 1 + 2N - 2N^{\text{ifo}}_{\texttt{NaNs}}$ and $\mathcal{K} \leq 2N - 1$. The overall number of contiguous \texttt{NaNs} in Eq.\,\eqref{eq:approx_delay_x} is given by 
\begin{equation}
N^{\eta}_{\texttt{NaNs}} = \underbrace{N^{\text{ifo}}_{\texttt{NaNs}}}_{\text{Telemetry}}  +  \underbrace{\frac{\mathcal{K} + 1}{2} + N}_{\text{Gap Augmentation}}\,.\label{eq:gap_augmentation_intermediate_regime_eta}
\end{equation}
Our third and final regime for the gap augmentation occurs due to a complete merging between the two sets $\mathcal{S}^{\delay_{ij}(\text{ifo})}_{\texttt{NaNs}}$ and $\mathcal{S}^{\text{ifo}}_{\texttt{NaNs}}$ and will be the focus of the next section.

\subsection*{Regime 3: Complete Merging}

We now focus on the final (and simplest) case where the set of telemetry NaNs are contained within the set of delayed telemetry NaNs. In other words, when  $\delay_{ij}\text{ifo}_{ij}[ n\in \mathcal{S}^{\text{ifo}}_{\texttt{NaNs}}] = \texttt{NaN}$ with $\mathcal{S}^{\text{ifo}}_{\texttt{NaNs}} \subseteq \mathcal{S}^{\delay_{ij}(\text{ifo})}_{\texttt{NaNs}}$.

The statement that $\mathcal{S}^{\text{ifo}}_{\texttt{NaNs}} \subseteq \mathcal{S}^{\delay_{ij}(\text{ifo})}_{\texttt{NaNs}}$ means that the gap segment in the telemetry variables is encapsulated within the missing data segment. From Eq.\,\eqref{eq:number_nans_delayed_set}, we understand that the \emph{total number} of \texttt{NaNs} will be $\mathcal{K} + N^{\text{tel}}_{\texttt{NaNs}}$. This means that the overall gap augmentation for $\mathcal{K} > 2N - 1$  must be
\begin{equation}
N^{\text{tel}\rightarrow \eta}_{\texttt{NaNs}} = \mathcal{K} \,. \label{eq:eta_nans_widening_last_regime}
\end{equation}
and overall number of missing data given by
\begin{equation}
N^{\eta}_{\texttt{NaNs}} = \underbrace{N^{\text{ifo}}_{\texttt{NaNs}
}}_{\text{telemetry}} + \underbrace{\mathcal{K}}_{\text{Gap Augmentation}} \,. \label{eq:eta_nans_last_regime}
\end{equation}
Putting all of the pieces together with Gap Augmentations Eq.\,\eqref{eq:gap_augmentation_eta_regime_1}, Eq.\,\eqref{eq:gap_augmentation_intermediate_regime_eta}, Eq.\,\eqref{eq:eta_nans_widening_last_regime} specific to the constraints Eq.\,\eqref{eq:first_transition_point} and Eq.\,\eqref{eq:second_transition_point} results in Eq.\,\eqref{eq:gap_augmentation_eta_analytical} defined in the main body of the text.

\section{Gap Augmentation with Nested Delays}\label{app:nested_delays_formulas}
The gap augmentation formula leading to Eq.\,\eqref{eq:gap_augmentation_eta_analytical} holds when considering single delay $\delay_{ij}$  on the ifo measurements. Deriving the gap augmentation where nested delays are utilized follows a near identical procedure discussed above. Consider, for example,
\begin{align}
y[n] &= x[n] + \delay_{i_0i_1\dots i_{m-1}}x[n] \label{eq:x_delay_equation_nested_appendix} \\  
&\approx x[n] + \sum_{k=1-p}^{p}x[n - N + k]
\label{eq:y_eta_equation_nested_delay}\,,
\end{align}
with integer part $N = \lfloor m\cdot \bar{L}/(c\Delta t)\rfloor$. Understand now that Eq.\,\eqref{eq:y_eta_equation_nested_delay} is in the same form as Eq.\,\eqref{eq:reduced_eta_appendix} only with a different integer part $N$ given by the sum of delays of individual links at specific time-stamps. We remark here that this simple expression is only valid for equal and constant arm-lengths. From near identical calculations that lead to Eq.\,\eqref{eq:gap_augmentation_eta_analytical}, we can compute the gap augmentation induced in Eq.\,\eqref{eq:y_eta_equation_nested_delay}.  
\begin{widetext}
\begin{equation}\label{eq:gap_augmentation_x_y_nested}
N^{x\rightarrow y}_{\texttt{NaNs}} = f(\mathcal{K}, N^{x}_{\texttt{NaNs}}, N) = \begin{cases}
    \mathcal{K} + N^{x}_{\texttt{NaNs}} & \text{if} \  1 < \mathcal{K} \leq 1 + 2N - 2N^{x}_{\texttt{NaNs}}  \\
    \frac{\mathcal{K} + 1}{2} + N  &1 + 2N - 2N^{x}_{\texttt{NaNs}}  \leq \mathcal{K} \leq 2N - 1 \\
    \mathcal{K} &  \mathcal{K} \geq 2N - 1\,.
\end{cases}
\end{equation}
\end{widetext}

For $N^{x}_{\texttt{NaNs}}$ the number of contiguous nans in the sequence of $x$. Notice that the notation outlined in Eq.\,\eqref{eq:gap_augmentation_x_y_nested} can be used to compute the gap augmentation in the $\eta_{ij}[n]$ variables, where $N^{\eta \rightarrow \text{ifo}}_{\texttt{NaNs}} = f(\mathcal{K}, N^{\text{ifo}}_{\texttt{NaNs}}, N)$. 

The factorized expressions for the first and second generation TDI variables, derived from Eq.\,\eqref{eq:a_variables}, Eq.\,\eqref{eq:r_variables} and Eq.\,\eqref{eq:q_variables}, are written in an identical form to \eqref{eq:y_eta_equation_nested_delay}. Equation \eqref{eq:gap_augmentation_x_y_nested} will be used extensively in Section \ref{subsec:gap_augmentation_TDI_variables}.

\section{Multiple Gaps in Telemetry Variables}\label{app:multiple_gaps_telemetry_variables}
For multiple gaps in the telemetric data streams, we can construct the full set of $G$ gaps. 
\begin{align}
\mathcal{S}^{\text{ifo}}_{\text{gaps}} &= \{(\mathcal{S}^{\text{ifo}}_{\texttt{NaNs}})^{(0)},(\mathcal{S}^{\text{ifo}}_{\texttt{NaNs}})^{(1)},\ldots,(\mathcal{S}^{\text{ifo}}_{\texttt{NaNs}})^{(G - 1)}\}\,, \nonumber \\
&= \left \{ (\mathcal{S}^{\text{ifo}}_{\texttt{NaNs}})^{(i)}\right \}_{i=0}^{G-1}
\end{align}
where each element of $(\mathcal{S}^{\text{ifo}}_{\texttt{NaNs}})^{(i)}\in \mathcal{S}^{\text{ifo}}_{\text{gaps}}$ is a gap consisting of a contiguous set of missing-data of the form identical to \eqref{eq:telemetry_nans_set}
\begin{equation}\label{eq:i_th_missing_data_set_telemetry}
(\mathcal{S}^{\text{ifo}}_{\texttt{NaNs}})^{(i)} = \{n^{(i)}_0, n^{(i)}_0 + 1, \ldots, n^{(i)}_0 + (N^{\text{ifo}}_{\texttt{NaNs}})^{(i)} - 1\}\,.
\end{equation}
Here $n^{(i)}_0$ stands for the first index of the $i$-th gap segment with $i\in\{0,\ldots,G-1\}$ for $G$ number of data gaps present. Each gap segment $(\mathcal{S}^{\text{ifo}}_{\texttt{NaNs}})^{(i)}$ contains at most $(N^{\text{ifo}}_{\texttt{NaNs}})^{(i)}$ and do not overlap, i.e., 
\begin{equation}\label{eq:overlapping_criteria_multiple_gaps}
\bigcap_{i = 0}^{G - 1}(\mathcal{S}^{\text{ifo}}_{\texttt{NaNs}})^{(i)}= \emptyset \,.
\end{equation}

We will now discuss how to compute the gap augmentation in the intermediary variables $\eta_{ij}[n]$ for the general case where we have arbitrary numbers of gapped segments $(G > 1)$ each with potentially different missing data durations $(N^{\text{ifo}}_{\texttt{NaNs}})^{(i)} \neq (N^{\text{ifo}}_{\texttt{NaNs}})^{(j)}$. 

For each gap segment representing missing data in the telemetry variables $(\mathcal{S}^{\text{ifo}}_\texttt{NaNs})^{(i)}$, we proceed in a similar fashion to Appendix \ref{app:gap_augmentation_eta}, Eq.~\eqref{app_eq:delayed_indices_telemetry}, by computing the first and last indices of the summation term in Eq.\,\eqref{eq:delayed_eta_using_y} where missing-data is identified for each gap segment $i\in\{0,\ldots,G-1\}$
\begin{align}
(n^{\delay_{ij}(\text{ifo})}_{\text{first}})^{(i)} &= n^{(i)}_0 + N - p \label{eq:mult_gaps_delayed_first}\\
(n^{\delay_{ij}(\text{ifo})}_{\text{last}})^{(i)} &= n^{(i)}_0 + (N^{\text{(ifo,i)}}_{\texttt{NaNs}} - 1) + N - (1-p)\,.\label{eq:mult_gaps_delayed_second}
\end{align}
With these indices we can then construct the masked data sets for $i\in\{0,\ldots,G-1\}$
\begin{align}\label{eq:delayed_gap_set}
\mathcal{S}^{\delay_{ij}(\text{ifo})}_{\text{gaps}} &= \left\{(S^{\delay_{ij}(\text{ifo})}_{\texttt{NaNs}})^{(i)}\right\}_{i=0}^{G-1}
\end{align}
with first \eqref{eq:mult_gaps_delayed_first} and last \eqref{eq:mult_gaps_delayed_second} indices of the missing data placement in the $i^{\text{th}}$ missing data segment $(\mathcal{S}^{\delay_{ij}(\text{ifo})}_{\texttt{NaNs}})^{(i)}$. To construct the overall gated segment in the intermediary variables $\eta_{ij}[n]$, corresponding to Eq.\,\eqref{eq:reduced_eta_appendix}, we merge the set $(\mathcal{S}^{\delay_{ij}(\text{ifo})}_{\texttt{NaNs}})^{(i)}$ (in other words, take the union) with the original telemetry set $(\mathcal{S}^{\text{ifo}}_{\texttt{NaNs}})^{(i)}$ giving the missing data placements in gap $i$
\begin{equation}
(\mathcal{S}^{\eta}_{\texttt{NaNs}})^{(i)} = (\mathcal{S}^{\delay_{ij}(\text{ifo})}_{\texttt{NaNs}})^{(i)}\cup (\mathcal{S}^{\text{ifo}}_{\texttt{NaNs}})^{(i)}
\end{equation}
We now arrive at the complete data set representing the indices of \texttt{NaNs} present in the intermediary variables
\begin{equation}\label{eq:eta_multiple_gap_set}
\mathcal{S}^{\eta}_{\text{gaps}} = \{ (\mathcal{S}^{\eta}_{\texttt{NaNs}})^{(0)}, (\mathcal{S}^{\eta}_{\texttt{NaNs}})^{(1)}, \ldots, (\mathcal{S}^{\eta}_{\texttt{NaNs}})^{(G-1)}\}\,.
\end{equation}
The overall gap augmentation is then calculated by computing the difference of missing data values between $\mathcal{S}^{\eta}_{\text{gaps}}$ and $\mathcal{S}^{\text{ifo}}_{\text{gaps}}$.

Repeated application of the procedure allows one to construct the gapped sets for the Michelson-like TDI variables, $\mathcal{S}^{X_1}_{\text{gaps}}$ and $\mathcal{S}^{X_2}_{\text{gaps}}$ for arbitrary numbers of gaps with varying durations. A very similar approach to Sec.\eqref{subsec:gap_augmentation_TDI_variables} is followed, where one computes the gapped data sets for the $a_{1,2}$, $r_{1,2}$ and $q_{1,2}$ variables defined by Equations \eqref{eq:a_variables}, \eqref{eq:r_variables} and \eqref{eq:q_variables}. The only considerable change is the overall integer delay that must be considered when moving from masked data sets $a_{1,2}$, $r_{1,2}$ and $q_{1,2}$. Due to the extra delays, we must modify the first and last \texttt{NaN} indices of the delayed data sets [\eqref{eq:mult_gaps_delayed_first}, \eqref{eq:mult_gaps_delayed_second}] to have $N = \lfloor 2\bar{L}/(c\Delta t)\rfloor$ when augmenting from $a_{1,2}\rightarrow r_{1,2}$ and $N = \lfloor 4\bar{L}/(c\Delta t)\rfloor$ when augmenting from $r_{1,2} \rightarrow q_{1,2}$. Following this procedure allows one to construct both first and second generation masked data sets for arbitrary families of gap types 

\begin{subequations}\label{app_eq:list_missing_data_sets_X1_X2}
\begin{align}
\mathcal{S}^{X_1}_{\text{gaps}} &= \{(\mathcal{S}^{X_1}_{\texttt{NaNs}})^{(0)},(\mathcal{S}^{X_1}_{\texttt{NaNs}})^{(1)}, \ldots,(\mathcal{S}^{X_1}_{\texttt{NaNs}})^{(G-1)}\}\,, \\
\mathcal{S}^{X_2}_{\text{gaps}} &= \{(\mathcal{S}^{X_2}_{\texttt{NaNs}})^{(0)},(\mathcal{S}^{X_2}_{\texttt{NaNs}})^{(1)} \ldots,(\mathcal{S}^{X_2}_{\texttt{NaNs}})^{(G-1)}\}\,.
\end{align}
\end{subequations}
where the overall gap augmentation and final total number of missing data elements can then be easily calculated.

The procedure above is a general method to compute the overall number of missing data through propagation of gated telemetry through TDI. We will use the formalism presented in this section when we analyse a full-scale four-year long LISA data set sampled at $4$\,Hz in Sec.\eqref{subsec:results_full_scale_TDI_simulation_gaps}. Such simulations are too memory intensive and expensive to be generated using LISA simulation software such as \texttt{lisainstrument} and \texttt{pyTDI}.

Our discussion in this appendix is general and is applicable to any family of gap types with any duration and rate. Our only requirement is that the missing data points are consistent across all IFO measurements.

Although the methods discussed here will always be cheaper than propagating corrupted telemetry data through \texttt{pyTDI}, it still can be quite expensive to compute. Data sets that are four years in length sampled at 4\,Hz are large to store and, although the operations discussed above are relatively simple, there is still a vast quantity of data that must be processed. A cheaper and still very accurate way to compute the gap augmentation is to simply compute the gap augmentation in the various TDI variables using the analytical formulas \eqref{eq:Widening_of_X1_X2_subequations}. Given the list of missing-data telemetry sets \eqref{eq:i_th_missing_data_set_telemetry}, one can compute Eq.\,\eqref{eq:Widening_of_X1_X2_subequations}  

and sum the total amount of missing-data predicted from each segment. 

This is a very cheap way to estimate the number of missing data points in the overall TDI expressions with the only caveat being a potential overestimation of the number of missing data values. Such scientific claims should then be understood as conservative. Making reference to Fig.~\ref{fig:TDI_Variables_X_numerical_plot}, we see that for a fixed order of interpolate $\mathcal{K} = 45$, the maximum loss of data is approximately $\sim 90$\,s. This means that if there are multiple (telemetry-based) gaps contained within a 90 second interval, then the cheap gap augmentation procedure outlined above will not account for the intersecting data gaps as a result of widened edges. Using this cheap method leads to a possibility that we will slightly overestimate the number of missing data elements, which is ultimately conservative at heart.

\bibliographystyle{apsrev4-2}
\bibliography{gremlins}

\end{document}